\documentclass[pra,twocolumn,showpacs]{revtex4-1}
\usepackage{bm, bbm, amsmath, amssymb, amsthm}
\usepackage[T1]{fontenc}
\usepackage{color}
\usepackage[latin9]{inputenc}
\usepackage{graphicx}
\usepackage{times}

\newcommand{\be}{\begin{eqnarray}}
\newcommand{\ee}{\end{eqnarray}}
\newcommand{\bee}{\begin{equation}}
\newcommand{\eee}{\end{equation}}
\newcommand{\one}{\mathbbm{1}}
\newcommand{\Tr}[1]{\mathrm{tr}\left[ {#1} \right]}
\newcommand{\pTr}[2]{\mathrm{tr}_{#1}\left[ {#2} \right]}
\newcommand{\ket}[1]{|{#1}\rangle}
\newcommand{\bra}[1]{\langle{#1}|}
\newcommand{\braket}[2]{\langle{#1}|{#2}\rangle}
\newcommand{\ketbrad}[1]{|{#1}\rangle\!\langle{#1}|}
\newcommand{\ketbra}[2]{|{#1}\rangle\!\langle{#2}|}

\newcommand{\eg}{\emph{e.g.~}}
\newcommand{\ie}{\emph{i.e.~}}

\begin{document}

\title{Tensor network methods with graph enhancement}

\author{R.\ H\"ubener$^{1,2,3}$, C.\ Kruszynska$^{3}$, L.\ Hartmann$^{3}$, W.\ D\"ur$^{3}$, M.\ B.\ Plenio$^{4,5}$, and J.\ Eisert$^{1,2}$}
\affiliation{
1 Dahlem Center for Complex Quantum Systems, Freie Universit{\"a}t
Berlin, 14195 Berlin, Germany\\
2 Institute f{\"u}r Physik und Astronomie, University of Potsdam, 
14476 Potsdam, Germany\\
3 Institut f\"ur Theoretische Physik, Universit\"at Innsbruck, 
A-6020 Innsbruck, Austria\\
4 Institut f{\"u}r Theoretische Physik, University of Ulm, D-89069 Ulm, Germany\\
5 QOLS, Blackett Laboratory, Imperial College London, SW7 2BW London, UK}

\date{\today}

\begin{abstract}
We present applications of the renormalization algorithm with graph enhancement (RAGE). 
This analysis extends the algorithms and applications given for approaches based on matrix product states introduced in [Phys.~Rev.~A {\bf 79}, 022317 (2009)] to other tensor-network states such as the tensor tree states (TTS) and projected entangled pair states (PEPS).
We investigate the suitability of the bare TTS to describe ground states, showing that the description of certain graph states and condensed matter models improves.
We investigate graph-enhanced tensor-network states, demonstrating that in some cases (disturbed graph states and for certain quantum circuits) the combination of weighted graph states with tensor tree states can greatly improve the accuracy of the description of ground states and time evolved states.
We comment on delineating the boundary of the classically efficiently simulatable states of quantum many-body systems.
\end{abstract}

\pacs{03.67.Mn, 03.65.Ud, 03.67.Lx, 02.70.-c}

\maketitle

\section{Introduction}\label{sec:intro}

Quantum many-body systems show interesting emerging properties whose underlying mechanisms are hard to grasp but fundamental for the 
understanding of technological applications as well as the conceptual foundations of physics.
This applies, as far as the former is concerned, for example, to properties of superconductivity 
or of rare-earth magnetic insulators; concerning the latter, quantum many-body theory has had significant impact to the
understanding of the relationship between physical descriptions at different scales.
To gain insight and access to such complex systems, it proved useful to identify certain variational classes of states that are simple to describe and to analyze (\eg numerically) and yet carry the essential characteristics of the investigated systems. Such efforts have been made and carried out successfully for one-dimensional systems with renormalization methods. Following early attempts of real-space renormalization \cite{W75}, the {\it density matrix renormalization group (DMRG)} \cite{Wh92,Wh93,S05} provides a framework for the proper identification and treatment of the effective low-energy sector of most one-dimensional models.
An intimate connection between the DMRG and {\it matrix product states (MPS)} \cite{FNW92,OR95} has been identified, in that it has been understood that DMRG is a variational method over MPS. Interestingly, quantum information theory and its framework has provided a deeper understanding of the essentials behind this approach \cite{VPC04,RO97}.

Following the renormalization idea, many extensions of the DMRG concept have been proposed and used successfully for numerical and analytical access to complex systems. For instance, the logarithmic corrections of critical one-dimensional systems can be described by the {\it multi-scale-renormalization ansatz (MERA)} \cite{Vi07}. Moreover, to overcome the one-dimensionality of the DMRG, {\it tensor-tree states (TTS)} \cite{SDV06,MS06,TEV09,MLNV10,SPV08}	
 have been introduced and successfully applied to systems governed by tree-like interaction graphs \cite{DRS02, Ot96, Fr97, LCP00} and others. More generally, an arbitrary geometry of the description is allowed by {\it projected entangled pair states (PEPS)} \cite{VC06b}, following the idea to describe the state of a quantum system by suitable projections applied to a highly entangled state.

Apart from the possibility to describe and to characterize a state, it is furthermore essential to be able to compute expectation values and other physical quantities in an efficient way. While some tensor network states allow for exact calculations of local quantities (MPS, TTS, MERA), for others only approximate solutions are known. E.g., while the states occurring in the MERA approach can be efficiently represented as PEPS, the former allow for exact efficient evaluations of local observables, while the latter generally do not \cite{BKE10}.

Besides ground states, also states that evolve in time are in the focus of classical simulation. This does not only apply to states undergoing time evolution under a local Hamiltonian; in the quantum information community, also states that appear in intermediate steps of a quantum computation are of interest. These states show entanglement features that are essentially beyond a one-dimensional renormalization ansatz. However, research in quantum information and computation does not only offer new problems for classical simulations, but also alternative ways for the efficient description of quantum states. 

The classification of quantum circuits, whose applications modify a given quantum system and its entanglement features in a controlled way \cite{BR01, RB01}, gives rise to an according classification of quantum states. Some of these states allow for efficient descriptions and evaluations, while not necessarily agreeing with a renormalization ansatz. An example of such states is provided by the so-called {\it stabilizer states} \cite{GoPhD}, the related {\it graph states} \cite{Hei05,Hei04} and {\it weighted graph states} (WGS) \cite{HCDB05,ABD06}, which are constructed by the application of controlled phase gates and local unitaries to an initially separable pure state. 

The lack of an \emph{ad hoc} choice of an underlying geometry in the application of gates provides these states with remarkable entanglement features, possibly complementary to those provided by renormalization procedures. Under this point of view -- and also to assess how useful they are for the description of physical realizations of quantum information and condensed matter systems
-- the WGS have been investigated with promising results, but the accuracy of the description of states could have been improved \cite{ABD06}.

In this work we will study in detail a recent proposal of the authors to combine renormalization methods with quantum circuits, \ie to combine tensor-network states and the WGS, resulting in the {\it renormalization algorithm with graph enhancement (RAGE)}, and so called {\it RAGE states} \cite{HKHDVEP09}. Specifically, we focus on the question how to improve the tensor-network state description of ground states and time-evolved quantum many-body systems by a combination with the WGS. Extending the results of our previous publication, we will use a wider class of tensor-network states in the present paper, encompassing MPS, TTS, and PEPS. We provide applicable algorithms to compute reduced density matrices and to update these variational states in optimization procedures efficiently.

We will first consider different tensor-network states such as MPS and TTS. As we intend to apply the RAGE states to 2D and higher dimensional systems and, generally, systems with interaction patterns that give rise to a high amount of entanglement, we first investigate the applicability of (bare) tensor-network states to this kind of system. An improvement compared to MPS is expected, as the more general TTS allow for a better reflection of the physical geometrical features of the systems to be described. We thereby extend prior results by \eg Shi \emph{et al.} \cite{SDV06} and Martín-Delgado \emph{et al.} \cite{DRS02} regarding TTS, as well as of Refs.\ \cite{Ot96, Fr97, LCP00, SDV06,MS06,TEV09,MLNV10,SPV08}. 
We present a comparative investigation of MPS, TTS (and also PEPS) using general analytic considerations as well as numerical simulations. We describe exactly and approximately the ground states of certain graph states, a 2D spin-glass toy model and a 1D modified Ising model with transverse field and a $1/r$ long-range interaction, respectively.
The investigations clearly demonstate that the TTS can offer an effective improvement over MPS for certain state classes \eg subsets of the graph states.  However, the numerical simulations indicate that the non-linear structure of TTS over the linear structure of MPS does not have the significant effect on the achieved accuracy in these more realistic condensed matter systems, as the improvement is not very strong -- such as in case of a 2D spin glass -- and also the negative impact of the broken symmetry of the TTS description is even stronger sometimes than the positive effect of a better connectivity within the tensor-network (1D long-range Ising). Hence, in some situations the tree-like structure of the TTS makes them favorable for the description of models with broken symmetries as well as higher dimensional interaction structure, but this is not always the case.

In the main part of the paper, we consider the combination of MPS, TTS and PEPS with WGS, and give details of algorithmic implementations. We use these states for numerical simulations of several models, concentrating on condensed matter physics (2D Ising, 2D Heisenberg model) as well as (random) quantum circuits. The results of these simulations indicate that a significant improvement of the accuracy of the description can be achieved in some cases. For instance, models that have graph states as a ground state typically feature a high amount of entanglement which is difficult to describe using MPS or similar essentially one-dimensional tensor product states alone. While graph states are naturally included in the WGS description, slightly disturbed versions of these states (for example, in a model with random local magnetic fields) typically are not. As we see in the simulations, a disturbance can be well described by the tensor product state combined with the undisturbed weighted graph-description. A prominent example of graph states is the {\it toric code state} \cite{Ki03}, whose disturbed case we describe with very good accuracy using the RAGE states, as will be shown. Even though, the RAGE description has its limits, as for example typical condensed matter systems like the ground state of the 2D Heisenberg model are apparently still to far away from the RAGE states to be approximated well with this set.

The paper is structured as follows. In Section~\ref{sec:survey} we will briefly describe weighted graph states and their properties as well as the tensor tree states used in subsequent investigations and augmentation procedures. In Section~\ref{sec:comparisonMPSTTS} we will compare the MPS and the TTS and give examples, analytically as well as numerically, comparing their suitability in the description of condensed matter systems with different geometry. In Section~\ref{sec:combining} we will describe how to combine the MPS with the WGS, as well as the TTS and the WGS, providing suitable algorithms adapted for each case. In Section~\ref{sec:appofRAGE} we will then give applications of the RAGE class in the search for ground states as well as the simulation of quantum circuits. Section~\ref{sec:summary} contains the summary and conclusions. The appendix contains analogous considerations for the PEPS, in which case no numerical simulations were performed, as well as a discussion of details of our numerical implementations.

\section{A short survey on tensor-network states and weighted graph states}\label{sec:survey}

In this section we will briefly describe the classes of quantum states used as fundamental objects in subsequent investigations and augmentation procedures. We describe the {\it matrix product states (MPS)}, {\it tensor tree states (TTS)}, {\it projected entangled pair states (PEPS)} and {\it weighted graph states (WGS)}. We give the applicable algorithms to compute reduced density matrices and to update these variational states in optimization procedures.

We will first focus on tensor-network states. These states obtain their name from a correspondence to a network, \ie a graph consisting of vertices and edges. The correspondence can be established as follows. A multi-partite state vector of qudits $\ket{\psi}$ can be written in a product state representation with local bases $\{ \ket{s_i} \}$, where $i=1,\dots ,N$ and $s_i=0,\dots ,D-1$, as
\be
\ket{\psi}=\sum_{s_1,\dots, s_N = 0}^{q-1} A_{s_1,\dots ,s_N}\ket{s_1 ,\dots ,s_N}.
\ee
The symbol $A$ is a tensor of rank $N$ with indices of dimensionality $q$. Hence there are $q^N$ complex numbers defining the tensor, and thus, the state. The tensor $A$ may now arise from a {\it contraction of other tensors}, \eg
\be
A_{s_1, s_2 ,s_3} = \sum_{\alpha, \beta, \gamma} A^{(1)}_{s_1, s_2, \alpha} A^{(2)}_{s_3, \alpha, \gamma} A^{(3)}_{\beta, \gamma},
\ee
where each tensor can be identified with a vertex in a graph. The Greek indices, which are summed over, define a connectivity relation of the tensors. They can hence be identified with edges in a graph where the tensors correspond to vertices. Usually, the local quantum systems are identified with vertices as well, characterized by only one edge connecting to them.

\subsection{Matrix product states}\label{sec:introMPS}

The class of matrix product states -- then referred as as {\it finitely correlated states} and expressed in the Heisenberg picture for infinite lattices -- was introduced in Ref.~\cite{FNW92}, in work that also provided a thorough analysis of their correlation and entanglement properties.
Later investigations revealed that DMRG numerical methods can indeed be viewed as variational methods based on MPS as ansatz states and the success of DMRG regarding the numerical simulation of one-dimensional quantum spin systems was understood from several perspectives, see \eg Refs.~\cite{VPC04,RO97}.
What is more, the good performance of DMRG is related to the insight that ground states of local Hamiltonians satisfy so-called {\it entanglement area laws} \cite{ECP10,LRV04,AEPW02, PEDC05, Wo06, CEPD06, WVHC08,SCV08a,H07} and are expected to contain in a sense little entanglement even for critical models.
For a recent comprehensive review, see Ref.\ \cite{ECP10}; for a discussion in the context of the present work, see Section \ref{sec:comparisonMPSTTS}. Let us start by giving a brief definition of MPS.

\subsubsection{Definition and notation}

There are two kinds of MPS, one having open and one having closed boundary conditions. Using a product basis  
$\{\ket{s_1,\dots, s_N}\}$, with $s_j= 0,1$, an $N$-spin or qubit MPS vector with closed boundary conditions is defined as
\bee
\ket{\psi_{\text{MPS, closed}}}= \sum_{s_1 ,\dots ,s_N=0}^{1} \Tr{A^{(1)}_{s_1}  \dots A^{(N)}_{s_N}}\ket{s_1 ,\dots ,s_N}.
\label{eq:MPS CBC}
\eee
In contrast, an $N$-spin MPS vector with open boundary conditions is defined as
\bee
\ket{\psi_{\text{MPS, open}}}=\sum_{s_1 ,\dots, s_N=0}^{1}a^{(1)}_{s_1}A^{(2)}_{s_2} \dots A^{(N-1)}_{s_{N-1}}a^{(N)}_{s_N}\ket{s_1 ,s_2,\ldots ,s_N}.
\label{eq:MPS OBC}
\eee
Here, the symbols $A^{(j)}_{s_j}$, with values $j=0,1,\dots ,N-1$, denote complex $D\times D$ matrices, and the symbols $a^{(1)}_{s_1}$ as well as $a^{(N)}_{s_N}$ denote complex $D$-dimensional row and column vectors, respectively.
For a discussion of the entanglement features of MPS, please see Section \ref{sec:comparisonMPSTTS}, where an analysis in comparison with the entanglement features of tensor tree states (TTS) is presented.

\subsubsection{Evaluation of observables in MPS}\label{sec:observable evaluation in MPS}

The following algorithms are applicable to the MPS but are neither new nor the most efficient implementations possible. They, however, form the basis for the algorithms in the corresponding RAGE states. 
Expressing a state as an MPS with a small value of $D$ allows for an efficient computation of reduced density matrices with small support. Consider a closed boundary matrix product state vector $\ket{\psi_{\text{MPS, closed}}}$, see Eq.~\eqref{eq:MPS CBC}, and let us define the transfer matrices
\be
E_{k,l}^{(j)}:=A_k^{(j)} \otimes (A_l^{(j)})^*,
\ee
where $*$ denotes complex conjugation of the elements of the matrix. 
The reduced state with support on sites in the subset $\mathcal{S} := \{m_1,\dots,m_{|\mathcal{S}|}\} \subset \{1,\dots,N\}$ with complement $\bar{\mathcal{S}}$ is then found to be
\bee
\begin{split}
	\rho_{\mathcal{S}}&=\pTr{\mathcal{S}}{\ketbrad{\psi_{\text{MPS, closed}}}} \\
	&= \sum_{{s_1,\ldots,s_N=0}\atop{r_1,\ldots,r_N=0}}^1 \Tr{E_{s_1,r_1}^{(1)} \ldots E_{s_N,r_N}^{(N)}} \\
	&\quad \times \pTr{\bar{\mathcal{S}}}{|s_1,\ldots ,s_N\rangle \langle r_1,\ldots , r_N|}\\
	&= \sum_{{s_{m_1},\ldots,s_{m_{|\mathcal{S}|}}=0} \atop{r_{m_1},\ldots,r_{m_{|\mathcal{S}|}}=0}}^1
	 \Tr{\prod_{n=1}^N T_n^{(n)}} \\ &\quad \times |	s_{m_1},\dots, s_{m_{|\mathcal{S}|}}\rangle \langle r_{m_1},\dots, r_{m_{|\mathcal{S}|}}|	
	\label{reducedMPSmatrix}
\end{split}
\eee
where
\be
T_n^{(n)}:=
\begin{cases}
\sum_{s_n} E_{s_n,s_n}^{(n)}, & n \in \bar{\mathcal{S}} \\
E_{s_n,r_n}^{(n)}, & n \in \mathcal{S}
\end{cases}.
\ee
For an $N$-qudit system with local physical dimension $q$, 
the effort to compute $\rho_{\mathcal{S}}$ scales as $O(|\mathcal{S}| D^5 q^{2 |\mathcal{S}|})$, as for each of the $q^{2 |\mathcal{S}|}$ elements in the reduced density matrix we have to multiply of the order of $|\mathcal{S}|$ matrices of dimension $D^2 \times D^2$, with an initial effort of $O(N D^5)$ to precompute the product of the matrices $T_n$ with support in $\bar{\mathcal{S}}$. As the number of matrix elements grows exponentially with the support of the reduced density matrix, only operators with small support, as the physically motivated type of Hamiltonian given above, are to be considered. 

\subsubsection{Variational methods in MPS}

Any MPS can be interpreted as a linear superposition of other MPS, where the linear superposition is controlled by the elements of one of its matrices that can be chosen arbitrarily. This dependence can be used to update matrix entries in order to optimize expectation values or overlaps. Consider writing the MPS in the form
\be
\ket{\psi_{\text{MPS, closed}}}=
\sum_{k,l,s_n} (A^{(n)}_{s_n})_{k,l} \ket{\mu(k,l,s_n)},
\ee
using another set of MPS defined by
\begin{multline}
\ket{\mu(k,l,s_n)} := \sum_{s_1, \dots, \hat{s}_n, \dots ,s_N} \Tr{A^{(1)}_{s_1} \dots D(k,l)^{(n)} \dots A^{(N)}_{s_N}}\\
\times \ket{s_1 ,\dots ,s_N},
\end{multline}
where
\bee
D(k,l)_{\tilde{k},\tilde{l}} = \delta_{\tilde{k},k}\delta_{\tilde{l},l}.
\label{eq:dummytensor}
\eee
This linear dependence of the MPS on its matrix entries and the reformulation of the original MPS into a superposition of other MPS makes it possible to efficiently compute a new representation of any operator with small support as a quadratic form \emph{over the entries of the vector $A$}. For example, a Hamiltonian $H$ is transformed into a matrix $\tilde{H}$, where the triples $k,l,r_n$ of indices are combined into one index $(k,l,r_n)$, etc.,
\be
\tilde{H}_{(k,l,r_n),(k',l',r'_n)}:=\bra{\mu(k,l,s_n)}H\ket{\mu(k',l',s'_n)}.
\label{eq:MPS_Helem}
\ee
Analogously, a new representation of $\one$ is obtained such that the energy can be written as a Rayleigh quotient
\be
\langle H \rangle = \frac{A^{\dagger} \cdot \tilde{H} \cdot A}{A^{\dagger} \cdot \tilde{\one} \cdot A},
\label{eq:expvalue}
\ee
where -- if the matrix $A$ belongs to site $n$ -- it can equally be regarded as a vector with elements 
\begin{equation}
A_{(k ,l ,s_n)} := (A^{(n)}_{s_n})_{k,l}, 
\end{equation}	
which we simply give the same symbol as the meaning is clear from the respective context. The extremalization of the expression in Eq.~\eqref{eq:expvalue} with respect to the entries in the vector $\alpha$ and hence in the matrices $A^{(n)}_{s_n}$ corresponding to a site $n$ can be achieved with linear algebra methods and essentially amounts to a generalized eigenvalue problem. A sequence of MPS, monotonously approaching an extremal point (\eg the ground state) can be computed by iterating the procedure over all values of $n$ repeatedly.
While this effectively amounts to solving a global optimization method by iterative local solutions, this method produces extraordinarily 
high accuracy in practice. Efficient algorithms for the case of open boundary MPS will be given in the next section, as these states are a special case of tensor tree states (TTS).

\subsection{Tensor tree states}\label{sec:TTS}

Recently Shi \emph{et al.}~\cite{SDV06} showed that open boundary MPS (see Eq.~\eqref{eq:MPS OBC}) are a special instance of a more general class of states, the {\it tensor tree states}, a description of states in terms of a tensor-network with tree structure, see also Refs.~\cite{MS06,Vi03,Vi04}. This class retains the favorable properties of MPS, in particular the ability to describe slightly entangled states, the possibility to update the description when such states evolve in real or imaginary time and the possibility to extract information from these states. We follow this approach and consider the description of states using a tree graph as the underlying network.

\subsubsection{Definition and notation}

Tensor tree states are tensor-network states where the network is a tree graph. In particular, (i) the leave nodes of the tree represent the local physical systems ($d$-level spins), (ii) a star-shaped graph with $k$ edges represents a tensor with $k$ indices, (iii) there are no loops in the graph, and (iv) edges in the graph represent indices shared by the corresponding tensors represented by the vertices and will be summed over. An $N$-qubit state in the computational basis is represented by a single tensor with $N$ indices, which can be obtained from any other tensor-network by performing all the different contractions.
\begin{figure}[t]
\includegraphics[width=0.65\columnwidth]{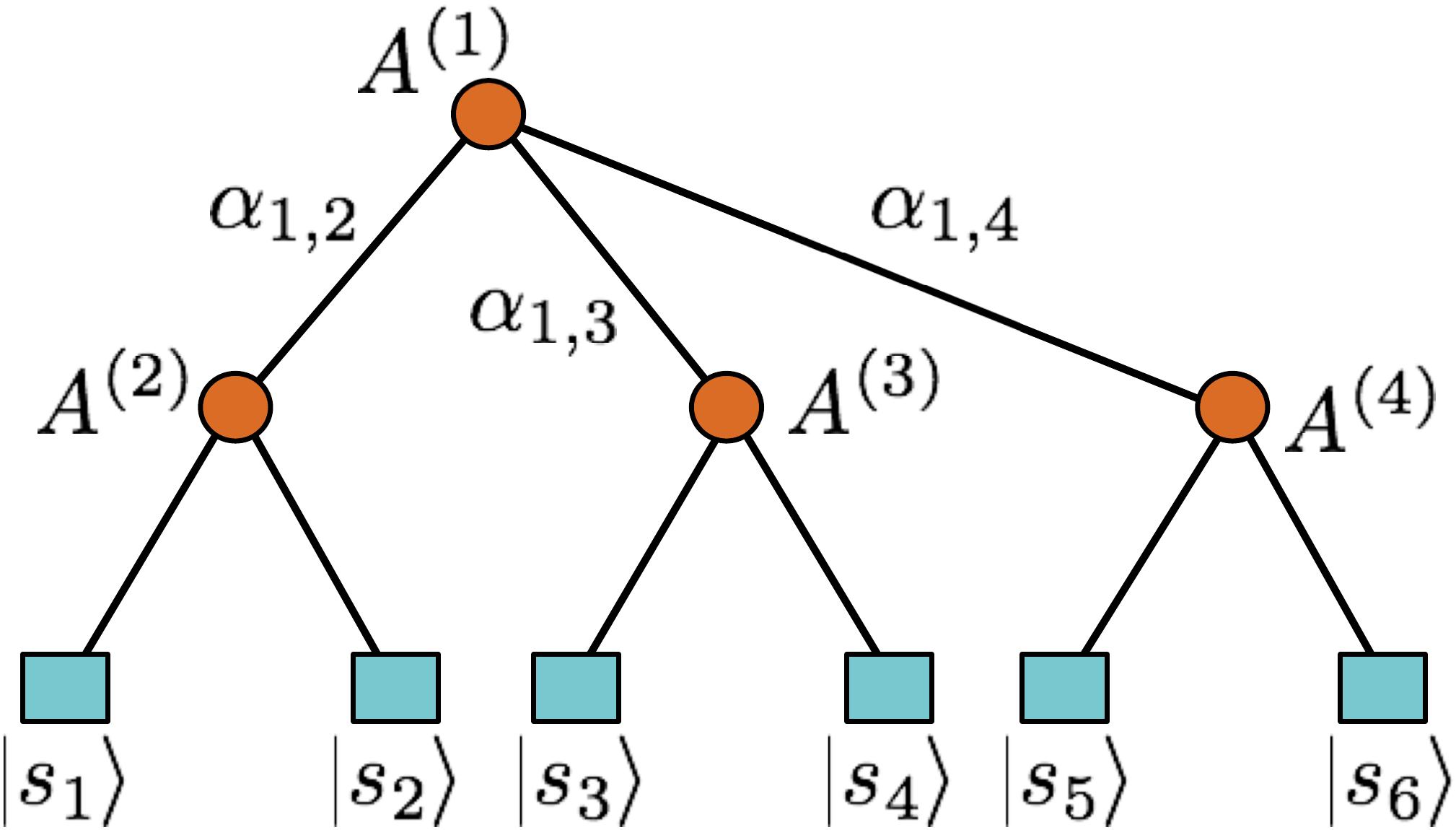}
\caption{(Color online) Subcubic tree representing a six-qubit quantum state.}
\label{fig:subcubictree}
\end{figure}
An important example of a tree is the so called subcubic tree, as depicted in Fig.~\ref{fig:subcubictree}. In a subcubic tree, the leaf-vertices have rank one, and all the other vertices rank three. The state vector represented by the tree in Fig.~\ref{fig:subcubictree} is correspondingly
\begin{multline}
\ket{\psi_{\text{TTS}}}=\sum_{\alpha_{1,2},\alpha_{1,3},\alpha_{1,4}=0}^{\chi-1}
\sum_{s_1,s_2,\dots,s_6=0}^{d-1}A^{(1)}_{\alpha_{1,2},\alpha_{1,3},\alpha_{1,4}}A^{(2) s_1 ,s_2}_{\alpha_{1,2}}\\
\times A^{(3) s_3 ,s_4}_{\alpha_{1,3}}A^{(4) s_5 ,s_6}_{\alpha_{1,4}}\ket{s_1,\ldots, s_6},
\label{eq:drawn TTS}
\end{multline}
or, if one singles out an arbitrary tensor, say $A^{(1)}$, the same state vector can be written as
\bee
\ket{\psi_{\text{TTS}}}= \sum_{\alpha_{1,2},\alpha_{1,3},\alpha_{1,4}=0}^{\chi-1}
A^{(1)}_{\alpha_{1,2},\alpha_{1,3},\alpha_{1,4}}\ket{\varphi^1_{\alpha_{1,2}}}\ket{\varphi^2_{\alpha_{1,3}}}\ket{\varphi^3_{\alpha_{1,4}}},
\eee
where $\ket{\varphi^i_{\alpha_{1,i}}}$ is
the part of the state with support on the spins connected to $A^{(1)}$ via the index $\alpha_{1,i}$, 
and similar for the other states of this kind. Note that these states are represented by 
subgraphs of the graph representing the whole TTS. 
\begin{figure}[t]
\includegraphics[width=0.93\columnwidth]{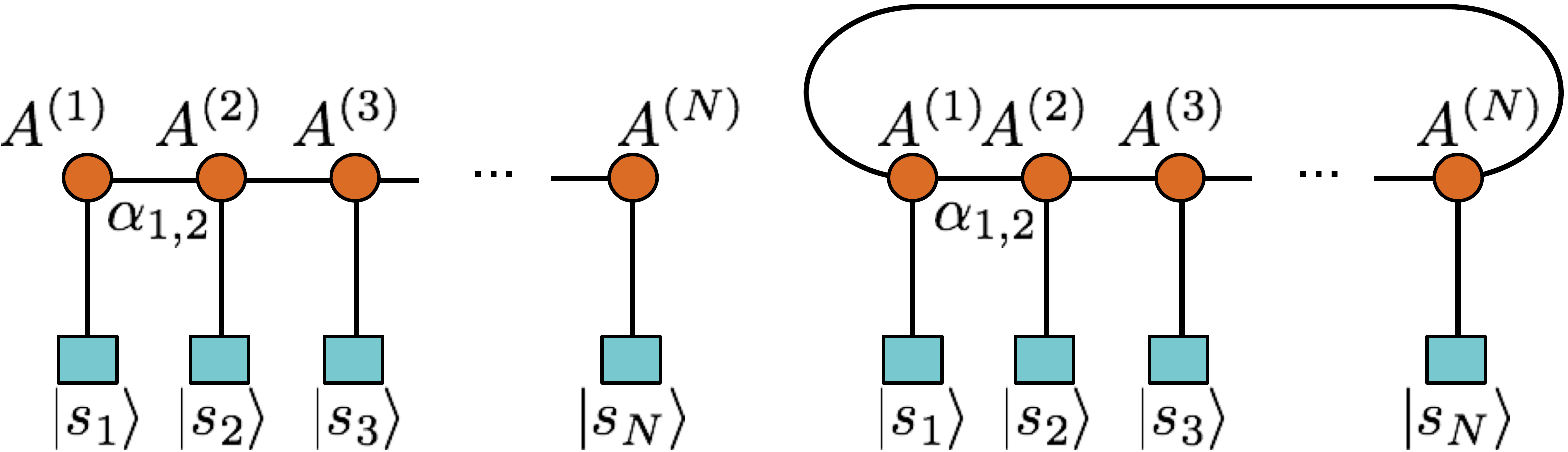}
\caption{(Color online) Tensor network representation of MPS with open and closed boundary conditions.}
\label{fig:MPStrees}
\end{figure} 

Consider the special case of an open-boundary MPS, so a state being defined by an 1D tensor network: 
Let us write the state corresponding to 
Fig.~\ref{fig:MPStrees}(a) and show that it is an MPS with open boundary conditions. 
The state vector corresponding to the given graph is
\begin{multline}
\ket{\psi_{\mathrm{MPS}}}= \sum_{\alpha_{1,2}\alpha_{2,3}\dots\alpha_{(N-1)N}=0}^{D-1} 
\sum_{s_1,\dots, s_N=0}^{d-1} A^{(1)s_1}_{\alpha_{1,2}} \dots \\
A^{(2)s_2}_{\alpha_{1,2} \alpha_{2,3}} A^{(N)s_N}_{\alpha_{(N-1),N}} \ket{s_1,\dots, s_N}.
\end{multline}
If we define the vectors
\begin{equation}
	(a^{(1)}_{s_1})_{k} = (A^{(1)s_1})_{k},\,\, 
 (a^{(N)}_{s_N})_{k} = (A^{(N)s_N})_{k},
\end{equation}  
and the matrices (tensors with two indices) 
\begin{equation}
	(A^{(i)}_{s_i})_{k,l}=(A^{(i)s_i})_{k,l},
\end{equation}
for 
$i\in\{2,3,\dots,N-1\}$, each coefficient $\braket{s_1, \dots ,s_N}{\psi_{\mathrm{MPS}}}$ of the 
state vector is given by the product of matrices $A^{(2)}_{s_2}A^{(3)}_{s_3}\dots A^{(N-1)}_{s_{N-1}}$ 
multiplied by the row vector $a^{(1)}_{s_1}$ from the left and by the column vector 
$a^{(N)}_{s_N}$ from the right. We obtain
\begin{equation}
\ket{\psi_{\mathrm{MPS}}}=\sum_{s_1, \dots , s_N=0}^{d-1}a^{(1)}_{s_1}A^{(2)}_{s_2}\dots a^{(N)}_{s_N}\ket{s_1, s_2,\dots,  s_N}, 
\end{equation}
which is an MPS with open boundary conditions as claimed.
For a discussion of the entanglement features of TTS, please see Sec.~\ref{sec:comparisonMPSTTS}, where an analysis in comparison with the entanglement features of MPS is given.

\subsubsection{Evaluation of observables}\label{sec:observable evaluation in TTN}

In this subsection we show how to evaluate product observables in an efficient way. Note that in particular the ability to compute sums of product observables allows us to compute the energy of a system described by a Hamiltonian that can be written as a sum of bilocal terms, \eg
\be\label{eq:general hamiltonian}
H=\sum_{\{a,b\}\in \mathcal{E}} H_{a,b}^{(a,b)},
\ee
where $\mathcal{E}$ denotes the set of pairs of spins connected by an interaction. Note that several interesting Hamiltonians such as the Ising or the Heisenberg Hamiltonian are of this form. Concerning our notation, as in the example Hamiltonian above, we will write a local operator $O$, acting on a multi-particle system with support on, say, site $a$ as $O^{(a)}:=\one \otimes \cdots \otimes O \otimes \cdots \otimes \one$. Correspondingly, 
we will write two-local operators as $O^{(a,b)}$ etc.

Let $\ket{\psi_{\text{TTS}}}$ be an $N$-qubit TTS, and $O=O_1\otimes O_2\otimes\dots \otimes O_N$ a product observable. We want to compute the expectation value $\bra{\psi_{\text{TTS}}}O\ket{\psi_{\text{TTS}}}$. The applicable algorithm is a recursive one, starting with an arbitrary edge $\alpha$ that is not connected to a leave. The edge defines a natural bi-partition of the system into the parts $L$ and $R$, corresponding to the subtrees connected by the edge. Following the definition of the TTS, we can write the expectation value with respect to the index $\alpha$ as
\be\label{eq:evaluate observable, step 1}
\bra{\psi_{\text{TTS}}}O\ket{\psi_{\text{TTS}}}=\sum_{\alpha\alpha'}\bra{ \varphi^L_{\alpha'}}O^L\ket{\varphi^L_{\alpha}}\bra{ \varphi^R_{\alpha'}}O^R\ket{\varphi^R_{\alpha}},
\ee 
where $O^{L}$ ($O^{R}$) is the part of the observable acting on $\ket{\varphi_{\alpha}^L}$ ($\ket{\varphi_{\alpha}^R}$). Let us write the left bracket explicitly, and let $A^L$ be the tensor connected to edge $\alpha$ and belonging to the subtree $L$. Using this definition,
we obtain $\ket{\varphi^L_{\alpha}}=\sum_{\beta_1,\beta_2}A^L_{\alpha,\beta_1,\beta_2}\ket{\varphi^1_{\beta_1}}
\ket{\varphi^2_{\beta_2}}$. Hence a recursion arises via the relation
\begin{multline}
\label{eq:phiaOphia'}
\bra{ \varphi^L_{\alpha'}}O^L\ket{\varphi^L_{\alpha}} = \sum_{\beta_1,\beta_2,\beta'_1,\beta'_2} A^{L}_{\alpha,\beta_1,\beta_2}
A^{L *}_{\alpha',\beta'_1,\beta'_2}\\
\bra{\varphi^1_{\beta'_1}}O^{L_1}\ket{\varphi^1_{\beta_1}}
\bra{\varphi^2_{\beta'_2}}O^{L_2}\ket{\varphi^2_{\beta_2}},
\end{multline}
which -- in a straightforwardly generalized manner -- is valid for all consecutive levels in the tree. The recursion is terminated if an expression of the kind $\bra{\varphi^2_{\beta'_2}}O^{L_2}\ket{\varphi^2_{\beta_2}}$ amounts to the expectation value of a local system, \ie if $\ket{\varphi^2_{\beta_2}}=\ket{s_n}$ for some local system $\ket{s_n}$.

The computational effort of this contraction is of order $O(T D^6)$ where $T$ is the number of tensors, which is of the order $O(N)$, provided that subtree expectation values are cached and reused in the recursion.

\subsubsection{Variational methods}\label{sec:optim in a TTNS}

The treatment of TTS is very similar to the treatment of MPS. The energy can be expressed analogously as a Rayleigh quotient, see Eq.\ \ref{eq:expvalue}. However the matrix elements of $\tilde{H}$ are given by an expression adapted to the slightly different tensor network structure. Choosing an arbitrary tensor $A$ and writing
\be\label{eq:TTS with T singled out}
\ket{\psi_{\text{TTS}}}=\sum_{\alpha,\beta,\gamma}A_{\alpha,\beta,\gamma}\ket{\varphi^1_{\alpha}}\ket{\varphi^2_{\beta}}\ket{\varphi^3_{\gamma}},
\ee
accordingly, we now have
\begin{equation}
\tilde{H}_{(\alpha',\beta',\gamma')(\alpha ,\beta ,\gamma)} := \bra{\varphi^1_{\alpha'}}\bra{\varphi^2_{\beta'}}\bra{\varphi^3_{\gamma'}}H\ket{\varphi^1_{\alpha}}\ket{\varphi^2_{\beta}}\ket{\varphi^3_{\gamma}},
\label{eq:effective matrices for TTS}
\end{equation}
and similarly for $\tilde{\one}$, see also Appendix \ref{sec:orthonormalization}.

\subsection{Weighted graph states}

{\it Weighted graph states (WGS)} \cite{HCDB05,ABD06} are a generalization of graph states \cite{Hei04,Hei05}. Both are multi-particle states derived from a graph whose vertices are identified with spins and whose edges correspond to (controlled) phase gates.

\subsubsection{Definition and notation}

A graph state corresponds to a graph $G=(V,E)$ of vertices $V$ and edges $E$. The graph implies a construction rule for a quantum state: We start with a set of $N=|V|$ quantum sites, one quantum site corresponding to exactly one vertex in the graph $G$. Initially, the quantum sites are prepared to be in the product state vector $\ket{+}^{\otimes N}$, and then, following the connectivity in the graph, a phase gate $\text{diag}(1,1,1,-1)$ is applied to two quantum sites whenever the corresponding vertices are connected by an edge. A prominent example of such graph states is the 2D cluster state~\cite{BR01}, corresponding to a two-dimensional rectangular lattice with phase gates acting on nearest neighbors.

A generalization of this contruction is given by the \emph{weighted} graph states, where each phase gate can be a unique controlled phase gate. Accordingly, a weighted graph state vector is given by 
\be
\ket{\psi_G}=\left(\prod_{\{a,b\}\in E}\Lambda Z^{(a,b)}_{\varphi_{a,b}}\right)\ket{+}^{\otimes N},
\label{eq:weighted graph states: interaction picture}
\ee
where 
\be\label{eq:W phase gate}
\Lambda Z_{\varphi_{a,b}}=\text{diag}\left(1,1,1,e^{i \pi \varphi_{a,b} /2})\right.
.
\ee
Note the factor $1/2$ in the exponent, corresponding to a twofold application on each gate each pair, as we do not limit the application to pairs $a<b$, which is often found in the literature.

A (weighted) graph can be described by a matrix, its adjacency matrix with entries
\be
\left(\Gamma_G\right)_{a,b} =
\begin{cases}
 \varphi_{a,b}&\textrm{if $a$ and $b$ are connected by $\varphi_{a,b}$}\\
 0& \textrm{otherwise}
 \end{cases},
\ee
where in a non-weighted graph state, the angles are given by either $0$ or $1$.  Both the rows and the columns of the matrix correspond one-to-one to the vertices in the graph -- and hence to the sites of the quantum many-body system.

Non-weighted graph states have another important interpretation. They are the common eigenstate with eigenvalue $1$ for the set of commuting operators given by
\be
\label{Koperators}
K_a = \sigma_x^a \sigma_z^{N_a}:= \sigma_x^a \prod_{b \in N_a} \sigma_z^b,
\ee
where $a$ denotes a vertex and $N_a$ is the set of vertices connected to vertex $a$ via an edge. The operators $K_a$ form a complete set of commuting operators, comparable to the complete set of commuting observables characterizing the pure states of other quantum mechanical 
systems.

\subsubsection{Entanglement features}

Unlike MPS and TTS, the WGS bear no \emph{ad hoc} inherent geometrical relations and can hence be used to describe states arising from any interaction pattern. Hence each pair can be treated on equal footing and no preference of any geometric relation between individual sites is assumed nor implied. 

For example, the WGS may possess an arbitrarily high amount of entanglement insofar as the entanglement between any block of $N_A$ particles and the rest of the system may scale with the volume of the block \ie with the number $N_A$ of contained particles. In particular, the WGS are able to fulfill an area law \cite{ECP10}, and thus weighted graph states satisfy a condition which is necessary to approximate the ground state of many lattice spin systems. The set of weighted graph states contains the two-dimensional cluster state, which was shown to maximize the amount of entanglement under several different points-of-view (see, \eg Ref.~\cite{VDMB07}). Moreover, the correlation length within the WGS may diverge \cite{DHHLB05}.

\section{A comparison of matrix product states and tensor tree states}\label{sec:comparisonMPSTTS}

\subsection{Discussion}
The RAGE states are a composition of tensor-network and weighted graph states. Before investigating the features of the resulting composite class of states, its features and applicability, it is useful to first analyze both kinds of states individually and see what their respective characteristics are. Thus, in this section, we first investigate the applicability of (bare) tensor-network states such as MPS and TTS to describe systems with particular entanglement characteristics. As we are interested in 2D and 3D systems and, generally, systems with interaction patterns that give rise to a high amount of entanglement, we concentrate on this kind of system.

Since MPS are at the core of a very successful numerical method, the density matrix renormalization group (DMRG) method, it is an intriguing question whether the use of TTS, which are a superset of the MPS, may offer advantages for numerical methods. More precisely, we may ask: Given a Hamiltonian, is there a TTS that can approximate the ground state of this Hamiltonian more accurately than an MPS, while using the same number of parameters? And, moreover, when the system size grows, does the number of parameters in the TTS description grow more slowly compared to an MPS description of the state, for a desired accuracy? Unsurprisingly, these questions are related to the entanglement of the system and its scaling. In this spirit we will first construct a specific example where general TTS seem to be the better choice. This example should serve as a motivation why we include TTS into the RAGE class later and not merely MPS. 

As shown in Ref.~\cite{FNW92}, the Schmidt rank $\chi$ of a state is the quantity that tells us how efficiently MPS can describe this state. Any finitely sized state can be written as an MPS, provided that the matrix dimension $D$, limiting the Schmidt rank $\chi$ of the MPS strictly, is sufficiently large. 
Hence, a small Schmidt rank asserts a good description with small matrices. While it gives the right picture, this criterion is very strict, and can be relaxed by a consideration of the Renyi entropy instead. For a refined discussion, see Ref.\ \cite{SCV08a}. Characterizing the entanglement properties of systems to be described is hence important to estimate the possible fidelity. Important is the growth of the entanglement with the system size. The fact that an \emph{area law} holds means that the Renyi-entropy of the reduced state of a subset of sites is proportional to its surface area. Such an area law has been proven to govern the ground states of many interesting physical systems, see Refs.~\cite{ECP10,LRV04,AEPW02, PEDC05, Wo06, CEPD06, WVHC08}. However, an area law neither holds for (i) ground states of interacting systems with interactions that are too slowly decaying with distance, 
nor for (ii) ground states of critical systems, where the entropy of entanglement generally does not saturate but grows logarithmically with the system size at least in 1D systems \cite{Ca86}, nor for (iii) some time-evolved systems: If one allows the time to grow with the block size that is being considered \cite{SWVC08,CDEO08}. Otherwise, an entanglement area law holds for all times \cite{EO06,BHV06}.

In a TTS description, we have the following situation. A quick glance on Eqns.~\eqref{eq:phin} and \eqref{eq:phinplusone} in Appendix
\ref{sec:orthonormalization}
shows that, in case the two subsystems are connected by a single edge, the Schmidt rank is limited by the value of the index dimension $D$ of the edge. Hence for a subsystem $A$ connected to the remainder of the system by a single edge in the graph the entropy of entanglement of the reduced state is limited by $S(\rho_A)=\log_2 D$. More generally, each edge connecting two subsystems represents one ebit (carrying $D$-levels) of possible entanglement between the systems. Although there are bi-partitions that imply cuts through several edges in the graph and hence have a possibly larger entanglement, the maximum entropy of $\log_2 D$ between bi-partitions along an edge renders the TTS (as the MPS) inefficient for the simulation of ground states of spin systems whose interaction graphs are essentially two- or higher dimensional.

\subsection{Analysis of an advantageous TTS description}

For a more thorough investigation of the TTS description, the Schmidt rank must be replaced by another measure known from graph theory, the Schmidt rank width $\chi_{\text{wd}}$. Its definition implies a minimization over all possible trees $T$, which are describing the system \cite{VDB07}
\be
\chi_{\text{wd}}(\ket{\psi}):=\min_{T}\max_{e\in T}\log_2\chi_{A_T^e,B_T^e}(\ket{\psi}),
\ee
where $\chi_{A_T^e,B_T^e}$ is the Schmidt rank of the bi-partition one obtains by deleting edge $e$ from the tree $T$. Another measure is the entanglement width $E_\text{wd}$ defined similarly, except that $\log_2\chi_{A_T^e,B_T^e}$ is replaced by the bi-partite entanglement entropy $E_{A_T^e,B_T^e}$. In general the relation 
\begin{equation}
	\chi_\text{wd}\geq E_\text{wd} 
\end{equation}
holds \cite{VDB07}.

While a favorable TTS description possibly exists, the tree that actually achieves the task is hard to compute. For graph states, however, efficient algorithms are known, see \eg Ref.~\cite{OPhD,VDB07}, where the most efficient tree (up to a constant factor overhead) can be derived from the stabilizer description of the state. In this light is possible to construct examples where TTS are better suited than MPS. One instance is the following interaction model. 
Any non-weighted graph state has the property to be the common eigenstate with eigenvalue $1$ for the all operators
\be
\label{eq:Koperators}
K_a = \sigma_x^a \sigma_z^{N_a}:= \sigma_x^a \prod_{b \in N_a} \sigma_z^b,
\ee
where $a$ denotes a vertex and $N_a$ is the set of vertices connected to vertex $a$ via an edge. Hence, a Hamiltonian with local interaction terms of the type $-\sigma_{x}^{(a)}\prod_{b\in N_{a}}\sigma_{z}^{(b)}$ as above has a ground state which is a graph state. We consider an interaction like above taking place on a tree graph, with arbitrary interaction strengths, and possibly also additional magnetic field terms. Without the latter, the ground state of this system turns out to be a graph state, see Ref.~\cite{Hei04,Hei05}, where the corresponding graph is given by the tree structure specifying the interaction pattern. As shown in Ref.~\cite{VDB07}, such a graph state can be described by a TTS rank $2$ only, independent of the number of particles involved. Consequently, there exists a TTS description in which the tensor dimensions do not have to grow with the system size but can stay constant. Since one needs $O(N)$ tensors to describe the state -- independently of the geometry of the tree underlying the TTS description -- the total number of parameters, as well as the effort to compute expectation values in these states, scales as $O(N)$.

On the other hand, one can show that a description in terms of a MPS requires matrix dimensions of up to $\chi_{{\rm MPS}}\propto N^{1/2}$, which is \emph{increasing in an unbounded fashion} with system size, although only polynomially. We would like to give an intuitive argument at this point, without formal proof, to rule out the possibility of an efficient MPS description. To do so, we give an intelligent guess about a good MPS tree and argue why it is already the optimal one.
\begin{figure}[t]
\includegraphics[width=0.8\columnwidth]{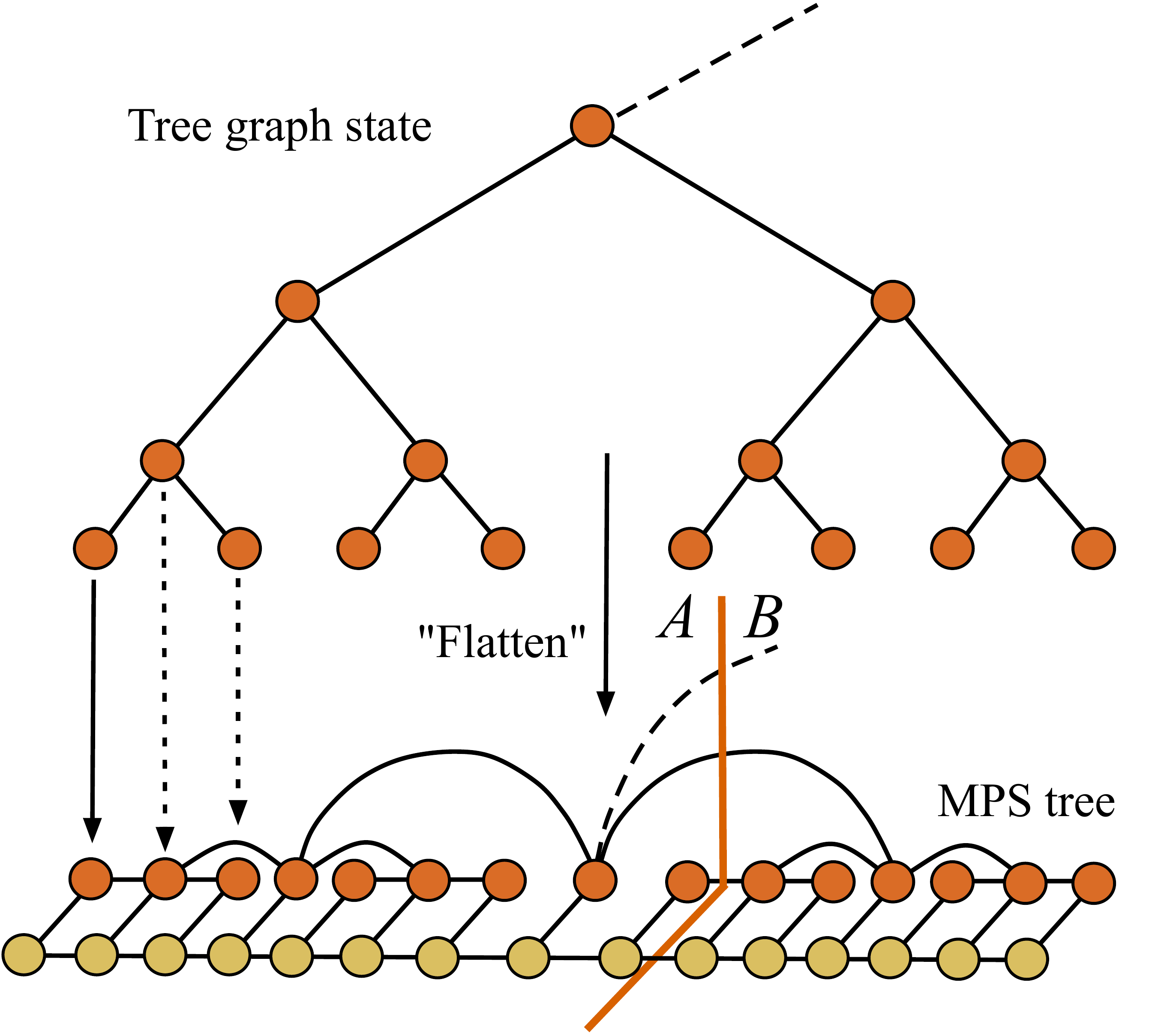}
\caption[]{(Color online) Tree graph state (top part) and specific arrangement (or labeling) of the qubits (bottom part) for an MPS description of the state. Qubits of the graph state and graph state interactions are shown in dark colors, while the MPS tree and points representing matrices are shown in light colors. One bi-partitioning is shown. The entropy in this case, according to the rules described in the text, is $2$ ebits.}
\label{FlatteningOfTreeGraphState}
\end{figure}

We proceed as sketched in Fig.~\ref{FlatteningOfTreeGraphState}. The tree graph state is flattened \ie we label the qubits in the order in which they appear when we move each single one down to the base line in the top part of Fig.~\ref{FlatteningOfTreeGraphState}. The MPS tree used to describe the state is sketched in the third dimension in the bottom part of the figure. The dimensions of the MPS matrices are determined by the Schmidt rank or the entropy of the chain's bi-partitions that are of the type ``right block vs. left block''. One can easily read off the maximum entropy of such bi-partitions by counting the black arches (graph state interactions) which are being cut by a vertical line specifying the bi-partition. All except one arch emerging from some specific qubit must be ignored. In this way one finds that there are some bi-partitions with $\lceil d/2 \rceil$ ebits where $d$ is the depth of tree corresponding to our tree graph state. Since $N \approx 2^d$ this number translates into MPS matrix dimensions of size at least $N^{1/2}\times N^{1/2}$. Hence, for any such matrix we need $O(N)$ parameters. A counting argument shows that the number of matrices with this dimension also grows with $N$, specifically like $N^{1/2}$. Hence, the total number of parameters for an MPS description with our chosen tree scales at least like $N^{3/2}$ as compared to a linear scaling that a more general TTS can achieve.

To show that the chosen MPS tree is the optimal one we would have to verify that (i) no rearrangement of any subset of qubits would lead to smaller entropies and hence to more favorable matrix dimensions and, even if this was the case, that (ii) no rearrangement can lead to the situation that the number of matrices that need $O(N)$ parameters becomes a constant,\ie independent of $N$. Any rearrangement of the qubits, unless it is a trivial one due to symmetries of the tree, will usually increase the number of interaction arches that bi-partitioning vertical lines cut. There are non-trivial rearrangements that do not increase this number, but these cannot lead to the situation described in (ii) as this would require that the interaction arches pile up above few qubit pairs, which is incompatible with the tree structure of the graph state.

\subsection{Numerical simulations}

In this subsection, we give the results of numerical simulations performed with MPS and TTS. We choose spin models that are usually hard cases for an MPS description and feature long-range, or two-dimensional, or broken-symmetry interactions. We show that in some cases the TTS, which can be adapted to non-symmetric interaction patterns, perform better than the chain-like MPS. 

The models were selected as they might show features relevant for the description of condensed matter systems while they are at the same time expected to be well suited for a tree description. For some very specific interaction models, as in the case of dendrimers
(this is a different kind of tree structure where all tensors correspond to a physical particle), 
the highly adapted TTS description already proved to reflect the physical situation well and achieved an accordingly good approximate description of many physical features of the system.

\begin{figure}[t]
\includegraphics[width=0.9\columnwidth]{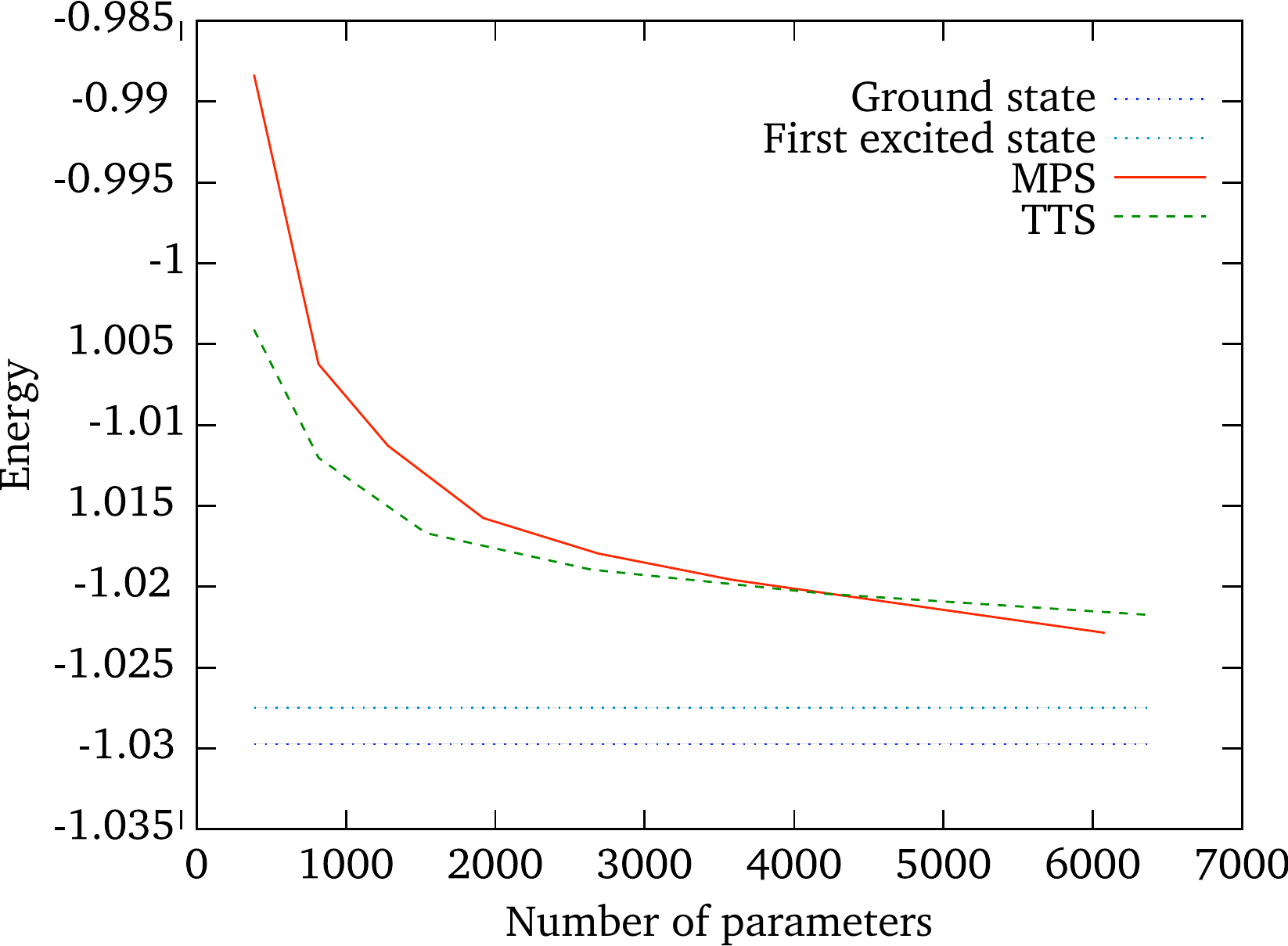}
\caption{(Color online) Spin-glass, 16 sites. Depicted is the achieved accuracy when approximating the ground state with MPS and TTS. We find that for a small number of parameters used in the description (\ie to signify the tensor entries), the sub-cubic tree describes a system better than the MPS. The situation is reversed for a large number of parameters.}
\label{Fig:spinglass}
\end{figure}

\subsubsection{A model with broken interaction symmetry}

The first numerically investigated case are spin glasses. As can be seen from this example, we find that for inhomogeneous systems the TTS are in some cases better suited to describe a state, as they can be adopted to the broken translational symmetry.
More precisely, we considered interactions of the anti-ferromagnetic Heisenberg type. We chose statistically distributed interaction strengths between next neighbors in a two-dimensional setting and with open boundary conditions. The model hence is described by the Hamiltonian
\be
H=\sum_{\langle i,j \rangle} J_{i,j} \left(\sigma_{x}^{(i)}\sigma_{x} ^{(j)}+\sigma_{y}^{(i)} \sigma_{y}^{(j)}+\sigma_{z}^{(i)}\sigma_{z}^{(j)}\right)
\ee
where $J_{i,j}$ follows a Gaussian distribution with mean $1$ and sigma $0.1$ and where $\langle i,j \rangle$ denotes nearest neighbors in a two-dimensional rectangular lattice. Spin sites that interact strongly are chosen to be represented by leaves that are connected closely in the tree. Since it is impossible to reflect the closeness relation consistently for all sites, a compromise has to be made (see below).
Generally, a priori considerations can be made of how certain properties of the model carry over to the entanglement structure of the state. These properties can be geometry-related (in real space or also in momentum space) or depend on details of the interaction itself. Besides considerations regarding the geometry of lattice models and the impact of neighborhood relations on correlations in short and long range interaction scenarios \cite{Ot96, LS03, BLGN10}, also typical electron correlations in certain models of quantum chemistry are subject to this kind of analysis \cite{RNW06, BLMR11}. An optimal choice of topological properties (correspondence between items in description reality, geometry) of the MPS or TTS graph can then be derived, taking into account \emph{e.g.} the von Neumann entropy.

The computational results for the search of the ground state energy of these systems are demonstrated in Fig.~\ref{Fig:spinglass}. We find that for a small number of parameters used in the description (\ie to signify the tensor entries), the TTS with sub-cubic tree structure describes a system better than the MPS. We remark that a similar behavior is also observed for certain regular systems (see below), however in these cases (as well as in the present example) the situation is reversed for a large number of parameters. 

This scaling behavior is due to the interplay of two concurrent effects. First, the number of edges between subsystems (connectivity) plays a role, as in MPS and TTS states correlations decay quickly over distances in the graph. Hence, in the case of nearest neighbor interactions, where close particles show often higher correlations than remote ones, the spin system is best described by reflecting the spatial closeness of physical sites by a closeness of the corresponding vertices in the graphical description. Thus, the connectivity favors the subcubic tree description, as the average number of edges between two sites is smaller there than in the flat tree description. 

Second, the bi-partite entanglement as measured by the Schmidt number grows merely as $D$, if the bi-partition of the state is performed by cutting an edge. Often, geometric effects are less influential than the scaling of the overall parameter count with $D$. In a tree, the number of parameters goes as $D^3$, whereas in an MPS description it grows as $D^2$. At some point, as the numerical simulations show, the different abilities of the different descriptions to adapt to the geometry of the model is overcome by the better scaling of the entanglement content in the simple MPS description.

\subsubsection{A chain with long-range interaction}

The second case we consider is a modified Ising model with transverse magnetic field and long-range interactions that decay as $1/r$. More precisely, we look at a chain governed by the Hamiltonian
\be
H=\sum_{{i,j=1}\atop{i \neq j}}^N \frac{1}{|i-j|} \sigma_z^{(i)} \sigma_z^{(j)} + \sum_{i=1}^N \sigma_x^{(i)}
\ee
with $N=24$.
\begin{figure}[t]
\label{Fig:modifiedIsing}
\includegraphics[width=0.9\columnwidth]{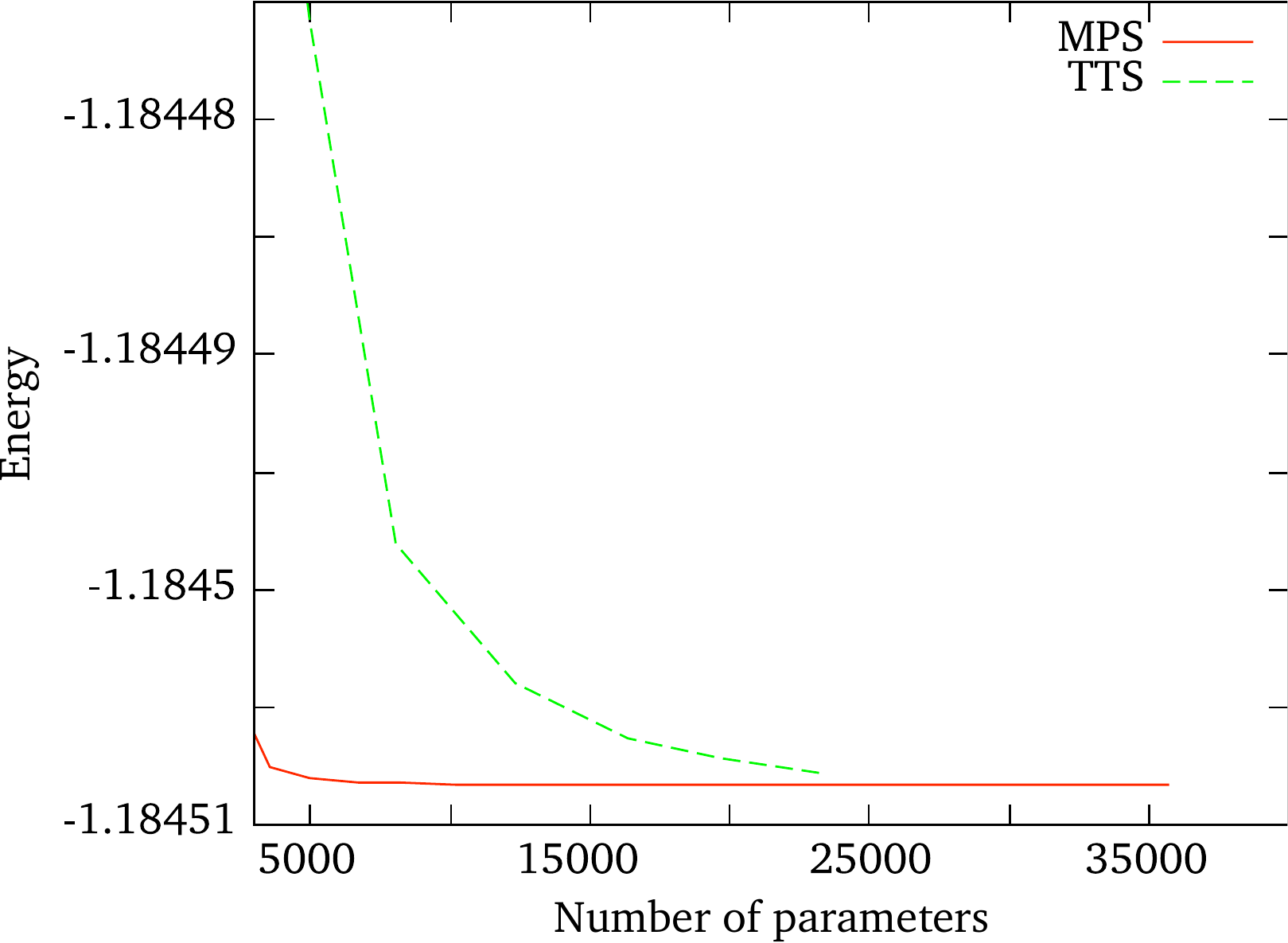}
\caption{(Color online) Modified Ising 1D, open chain, length 24, transverse field $B=1$, $\sigma_{z}\sigma_{z}-$coupling with strength $1/r$. Depicted is the achieved accuracy when approximating the ground state with MPS and TTS. Although the MPS are not suited to carry long range entanglement, they perform better than the sub-cubic tree, due to the fact that the tensor tree ansatz breaks the (translational) symmetries of the Hamiltonian explicitly.}
\end{figure}
The configuration chosen for the spin sites in the sub-cubic tree followed the idea developed above, \ie sites that interact strongly are chosen to be close in the tree. 

As a figure of merit for the description we computed the ground state energy, see Fig.~\ref{Fig:modifiedIsing}. Although in principle low dimensional MPS are not suited to carry long range entanglement, which is expected to occur in a system with such a Hamiltonian, in this model MPS always perform better than the sub-cubic tree. Note in this regard that the tensor tree ansatz breaks the (translational) symmetry of the Hamiltonian explicitly. The considerations made above in the context of the two-dimensional spin-glass apply also here.

For a related study of linear systems under long-range interaction as found in quantum chemistry models, see Ref.\ \cite{BLGN10}.

\section{Combining tensor-network states with weighted graph states}\label{sec:combining}

In this section, we want to consider in detail the renormalization algorithm with graph enhancement (RAGE). This improved renormalization algorithm corresponds to a variational state class, the RAGE states, which are a combination of tensor product states and WGS.
In the next sections we will give variants of RAGE states that correspond to certain special cases of tensor-network states. In the following we will discuss RAGE states based on MPS and TTS, as well as general tensor network states. The implementation is slightly different in each case, adapted to the underlying tensor network structure.

\subsection{RAGE states based on MPS}\label{MPSandWGS}

RAGE states based on MPS have been introduced in Ref.~\cite{HKHDVEP09}. We will now review the construction given in the same reference and provide additional details and discussions. For simplicity, clarity and to stress the connection to the algorithms given in the previous sections, we will restrict ourselves to the case where the spin-dimension is $q=2$ first, which is however easily generalized. Related results will be given in Sec.~\ref{sec:TTSandWGS}, in a modified form designed especially for the treatment of TTS and hence also open-boundary MPS, where moreover the applicable algorithms will be given explicitly for higher dimensional spin systems.

\subsubsection{Definition and notation}

We start from closed-boundary MPS of a quantum chain of length $N$, consisting of $2$-level systems, as in Sec.~\ref{sec:introMPS}.
\begin{equation}
|\psi(A)\rangle := \sum_{s_1,\ldots,s_N=0}^{1}
\Tr{A_{s_1}^{(1)}\ldots A_{s_N}^{(N)}}
|s_1,\ldots,s_N\rangle
\end{equation}
where the $A_{s_n}^{(n)}$ are complex $D\times D$ matrices. 

Now we consider the adjacency matrix $\varphi$ of a weighted simple graph with $\varphi_{k,l}\in [0,2\pi)$ and apply the corresponding phase gates $\Lambda Z(\varphi_{k,l})$ between the particles $k,l$ in the chain. Finally, we apply local rotations $V_j\in U(2)$, to arrive at the variational class of states defined by
\begin{multline}
|\psi(A,\varphi,V)\rangle:=
\prod_{j=1}^N V_j^{(j)} \prod_{k,l} \Lambda Z^{(k,l)}(\varphi_{k,l}) \\
\times \sum_{s_1, \ldots , s_N} \Tr{A_{s_1}^{(1)}\ldots A_{s_N}^{(N)}} |s_1, \ldots , s_N\rangle,
\label{RAGE}
\end{multline}
which then forms the basis of the {\it renormalization group algorithm with graph enhancement} based on MPS. For simplicity, and w.l.o.g., we will often set $V_j=\one$ subsequently.

\subsubsection{Relationship with MPS and WGS}

The RAGE states encompass both the MPS and WGS, and more. The MPS are included by definition, using $\varphi=0$ and $V_j=\one$. The inclusion of WGS is revealed by rewriting the expression from Ref.~\cite{ABD06}
\bee
\begin{split}
\ket{\psi}
=& \sum_m \alpha_m (\prod_{j=1}^N V_j^{(j)}) \\
&\quad \times \sum_{s_1,\ldots,s_N=0}^1  e^{-i  \mathbf{s}^T \varphi \mathbf{s} +  \mathbf{d}_m^T \mathbf{s}} \ket{s_1,\ldots ,s_N} \label{defwgraph} \\
=& (\prod_{j=1}^N V_j^{(j)} )(\prod_{k,l} \Lambda Z^{(k,l)}(\varphi_{k,l}))\\
&\quad \times \sum_m \alpha_m  \ket{\eta_{m,1}} \otimes \ldots \otimes \ket{\eta_{m,N}},
\end{split}
\eee
where $\mathbf{d}_m=(d_{m,1},\dots, d_{m,N})$, $\mathbf{s}=(s_1,\dots, s_N)$, $|\eta_{m,n}\rangle := \ket{0} + e^{d_{m,n}} \ket{1}$ and $\Lambda Z(\varphi_{m,n})$ are defined as above.

\subsubsection{Properties}

The RAGE states inherit and combine the properties of tensor product states and WGS. They have a polynomially sized description, where (in the present case) the MPS and the WGS part are fully determined by $O(N D^2)$ and $O(N^2)$ real parameters respectively.
Moreover, like the WGS, the RAGE states allow for long-ranged correlation and a volume law for the entanglement entropy. By having a collection of maximally entangled qubit pairs across a boundary, the von-Neumann entropy of a block of size $L$ can be taken to scale as $S(\rho_L) = O(L)$. Encompassing graph states, our class can hence maximize the entanglement entropy.

The given description of states allows for a \emph{manifest} translational invariance. Whenever the MPS part is translationally invariant, $\varphi$ is a cyclic matrix, and $V_j$ is the same for all $j$, the whole state $|\varphi\rangle$ is translationally invariant. There exist other translationally invariant states that do not have this simple form. The key feature is that there exists this natural subset of states for which translational invariance is guaranteed to be exactly fulfilled, while at the same time a volume law for block-wise entanglement is possible~\cite{DHHLB05,CHDB05}.
Finally, as MPS already form a complete set in Hilbert space (if one allows $D$ to scale as $O(2^N)$, one can represent any pure state in $(\mathbbm{C}^2)^{\otimes N}$) and this remains true for the RAGE set.

\subsubsection{Efficient computation of local properties and correlation functions}

The previous properties are all very natural and desirable, and especially a volume law for the entropy of reduced systems cannot be achieved efficiently with MPS alone. However, as will be shown, this does not prevent us from computing local properties and correlation functions efficiently.
To compute expectation values of observables with small support we use the relevant reduced density matrix $\rho_{\mathcal{S}}$, whose computation is efficient in the total size $N$ of the system ${\mathcal{S}}\subset \{1,\dots, N\}$ (see below). Controlled phase gates acting exclusively on qubits that are traced out make no contribution. We define 
\be
E^{(j)}_{k,l}:= A_{k}^{(j)}\otimes (A_{l}^{(j)})^*,
\ee
where $\ast$ denotes complex conjugation. The reduced density matrix $\rho_{\mathcal{S}}$ is then, by definition of the RAGE states, found to be
\begin{multline}
\rho_{\mathcal{S}} = \sum_{{s_1,\ldots,s_N=0}\atop{r_1,\ldots,r_N=0}}^1 \Tr{E_{s_1,r_1}^{(1)} \ldots E_{s_N,r_N}^{(N)}} \text{tr}_{\bar{\mathcal{S}}} [ (\prod_{k,l}\Lambda Z^{(k,l)}(\varphi_{k,l})) \\
\times \ketbra{s_1,\ldots ,s_N}{r_1,\ldots, r_N} (\prod_{k,l} \Lambda Z^{(k,l)\dagger}(\varphi_{k,l}))]
\end{multline}
The evaluation of the phase gates can be performed to obtain
\begin{multline}
\rho_{\mathcal{S}} = \sum_{{s_1,\ldots,s_N=0}\atop{r_1,\ldots,r_N=0}}^1\Tr{E_{s_1,r_1}^{(1)} \ldots E_{s_N,r_N}^{(N)}} \\
\times \ketbra{s_{m_1},\ldots ,s_{m_{|\mathcal{S}|}}}{r_{m_1}, \ldots , r_{m_{|\mathcal{S}|}}} (\prod_{k \in \bar{\mathcal{S}}} \delta_{s_k,r_k}) \\
\times (\prod_{k,l} e^{\pi i \varphi_{k,l}(\delta_{s_k,1}\delta_{s_l,1})/2}) (\prod_{k,l} e^{- \pi i \varphi_{k,l}(\delta_{r_k,1}\delta_{r_l,1})/2}).
\end{multline}

For the computation of the reduced density matrix, the effect of the phases is a mere modification of the transfer operators of the MPS by a phase factor, the phase depending on the matrix element in question. Thus, the evaluation of expectation values is (as in the case of MPS using the same ansatz for observable evaluation) performed using products of transfer operators associated with the single sites. The reduced state can then be written as
\begin{multline}
\rho_{\mathcal{S}}=\sum_{{s_{m_1},\dots, s_{m_{|\mathcal{S}|}}=0}\atop{r_{m_1},\ldots,r_{m_{|\mathcal{S}|}}=0}}^1 \Tr{\prod_{n=1}^N T^{(n)}} \\
\ketbra{s_{m_1},\dots, s_{m_{|\mathcal{S}|}}}{r_{m_1},\dots, r_{m_{|\mathcal{S}|}}},
\end{multline}
where now
\be
T^{(n)}&:=& 
\begin{cases}
\sum_{s_n=0}^1 E^{(n)}_{s_n s_n} \prod_{k \in \mathcal{S}} e^{2 i \Delta_{k,n}}, & n \in \bar{\mathcal{S}}\\
E^{(n)}_{s_n,r_n} \prod_{k \in \mathcal{S}} e^{i \Delta_{k,n}} ,& n \in \mathcal{S}
\end{cases},
\ee
with
\be
\Delta_{k,n} = \pi \varphi_{k,n} (\delta_{s_k,1}\delta_{s_n,1}-\delta_{r_k,1}\delta_{r_n,1}) /2.
\ee
Grouped in this way, the reduced density operator can indeed be evaluated efficiently. In fact, the computational effort for the reduced density matrix is merely $O(|\mathcal{S}| D^5 2^{2 |\mathcal{S}|})$, with an initial effort of $O(N D^5)$, as one has to multiply $N$ transfer matrices of dimension $D^2 \times D^2$ outside the support of $\mathcal{S}$, just as in the case of MPS. This procedure is inefficient in $|\mathcal{S}|$, with an exponential scaling effort. However, any Hamiltonian with two-body (possibly long-ranged) interactions can be treated efficiently term by term. Summing up, the evaluation of the expectation value of an operator that is the sum of $h$ terms with small support is of the order $O(h N D^5)$

Please note that the ansatz to compute expectation values via the reduced density matrix offered a natural and efficient numerical treatment by \emph{absorbing the WGS part into the MPS description}. As this ansatz scales exponentially in the size of the support of the observables, another one might be desirable. In Sec.~\ref{sec:TTSandWGS} we will shift the focus to another picture, where the WGS part is not absorbed into the state any more, but into intermediate auxiliary observables which will be used for the evaluation of the entries of the reduced density matrix. The complete step of absorbing the WGS part into the observable whose expectation value we want to compute will be made in Sec.~\ref{sec:extension}. One can take the point of view that these implementations are implementing different points of view on the modification performed by the graph enhancement. The border line is fuzzy, but the complete absorption of the WGS into the state -- as shown here -- and the complete absorption into the observables are in a way the extremal points of the implementation.

\subsubsection{Variational methods}

Apart from procedures for the efficient computation of reduced density matrices, and therefore expectation values, we need a variational principle to improve the trial states. We will focus on local variational approaches to approximate ground states by minimizing the energy, on the approximation of time evolution and on the simulation of quantum circuits. We note that the search for ground states is well known to be related to imaginary-time evolution.

The MPS part can be updated as shown in Sec.~\ref{sec:observable evaluation in MPS}. The expression $\bra{ \psi(A,\varphi,V)} H \ket{\psi(A,\varphi,V)}$ is, as for MPS, a quadratic form in each of the entries of the matrices $A^{(k)}_0$, $A^{(k)}_1$ for each site $k=1,\dots, N$. An optimal local update can therefore be found by means of solving generalized eigenvalue problems, with an effort of $O(D^6)$.

The optimization of the phases and local unitaries is possible in several ways. A self-evident method is to treat $E=\langle H \rangle$ as a function over all (polynomially many) parameters simultaneously and optimize it brute force with Nelder-Mead or gradient based methods, where one can take advantage of the possibility to analytically calculate the gradient. However, more systematic approaches are often desirable, as the energy possesses many local minima. Thus, after finding an initial local minimum with the just mentioned methods, we optimize local unitaries and phases one by one using a sweeping technique. Note that the phases and local unitaries cancel in the normalization term $\bra{ \psi(A,\varphi,V)} \one \ket{\psi(A,\varphi,V)}$ and hence do not enter the optimization problem.

The local rotations can be addressed by parameterizing single qubit rotations on spin $k$ with a normalized vector ${\bm x}_k \in \mathbbm{R}^4$ as 
\be
V_k=x_{k,0}\one+i(x_{k,1}\sigma_x-x_{k,2}\sigma_y+x_{k,3}\sigma_z).
\ee
Again, the local variation of ${\bm x}_k$, \eg for finding a minimum of the energy, amounts to a generalized eigenvalue problem in ${\bm x}_k$ for each site $k=1,\dots, N$.

The optimal phase gates between any pair of spins $j,k\in\{1,\dots,N\}$ can be optimized in another sweeping procedure over all phases, while keeping the tensorial part fixed. To do so, we first absorb the local unitaries into the Hamiltonian
\be \label{eq:phaseoptim one}
\tilde{H} := (\prod_{j=1}^N V_j^{(j)\dagger}) H (\prod_{k=1}^N V_k^{(k)}).
\ee
Then, to optimize the expectation value of $H$ with respect to a single phase, say $\varphi_{a,b}$, we reconsider the dependence between $\varphi_{a,b}$ and $\langle \tilde{H} \rangle$. For each local term $\tilde{H}_{c,d}^{(c,d)}$ we write 
\begin{widetext}
\begin{multline}
\bra{\psi_{\text{MPS}}}\prod_{k,l} \Lambda Z^{(k,l)}(-\varphi_{k,l}) \tilde{H}_{c,d}^{(c,d)} \prod_{k,l} \Lambda Z^{(k,l)}(\varphi_{k,l}) \ket{\psi_{\text{MPS}}}\\
=\bra{\psi_{\text{MPS}}} \prod_{{k,l}\atop{k,l \neq a,b}} \Lambda Z^{(k,l)}(-\varphi_{k,l}) \times \left(\Lambda Z^{(a,b)}(-\varphi_{a,b})) \tilde{H}_{c,d}^{(c,d)} \Lambda Z^{(a,b)}(\varphi_{a,b}))\right) \times \prod_{{k,l}\atop{k,l \neq a,b}} \Lambda Z^{(k,l)}(\varphi_{k,l}) \ket{\psi_{\text{MPS}}}.
\end{multline}
\end{widetext}
It is straightforward to show that for all operators $\tilde{H}_{c,d}^{(c,d)}$ the corresponding operator
\be \label{eq:operatortodissect}
\Lambda Z^{(a,b)}(-\varphi_{a,b})) \tilde{H}_{c,d}^{(c,d)} \Lambda Z^{(a,b)}(\varphi_{a,b}))
\ee
can be written as
\be \label{eq:operator sincosrelation}
\alpha + \beta \cos(\varphi_{a,b}) + \gamma \sin(\varphi_{a,b}),
\ee
with at most $3$-local hermitian operators $A,B,C$ (where $B,C$ are zero if $\{a,b\} \cap \{c,d\} = \emptyset$), and where the state vector 
\be
\prod_{{k,l}\atop{k,l \neq a,b}} \Lambda Z^{(k,l)}(\varphi_{k,l}) \ket{\psi_{\text{MPS}}}
\ee
is independent of $\varphi_{a,b}$. The optimization of the phases is thus tantamount to the optimization of an expression
\be
A + B \cos(\varphi_{a,b}) + \Gamma \sin(\varphi_{a,b}),
\label{eq:singlephase endform}
\ee
with efficiently computable real values $A, B \Gamma$. An element that is not present for MPS alone: One can make a choice whether one adapts an MPS part or the adjacency matrix for an identical change in the physical state. In practice, we have supplemented this procedure with an gradient-based global optimization, making use of the fact that the gradient can be explicitly computed.

We want to mention that, for practical applications, the optimization of the phases takes a lot of time. One reason is that many evaluations of expectation values in a RAGE state are needed to perform the Nelder-Mead optimization and also the individual updates of phase. One more problem is the bad convergence behavior of the phase optimization procedure, as the process usually comes to a halt on intermediate optimization levels for many optimization steps. A possible alternative to this approach would be a flow-inspired gradient method of optimizing over RAGE states \cite{DEO08}. In many cases, however, we will be interested in an approximation of states that have RAGE states as a natural description. These are, among others, noisy (meaning slightly disturbed) graph states, encompassing stabilizer states. In this case, good approximate phases for the noisy states can be found analytically. The numerical approximation can then be restricted to the tensorial part of the RAGE states, similar in performance to a bare MPS/TTS-approach, see also Sec.~\ref{sec:TTSandWGSupdates}.
To summarize, an update of $\ket{\psi(A,\varphi,V)}$ to minimize the energy corresponds to a sweeping over local variations, each of which is efficiently possible, with an effort of $O(M h N D^5)$ for $M$ sweeps.

\subsection{Time evolution and quantum algorithms}

A natural application of the renormalization with graph enhancement is the simulation of time evolution, where long range correlations arise. Instead of the full problem of a general time evolution we would like to consider the more specialized but nevertheless important and interesting issue of quantum algorithms. These will still exhibit long range correlations (otherwise they would be simulatable by standard methods) but have the advantage that they are already broken down into a discrete set of simple quantum operations. Indeed, every quantum algorithm may be decomposed into a sequence of general single qubit gates and controlled phase gates. Such a set of gates seems particularly well suited for treatment in the RAGE picture.

Let us consider the simulation approach in some more detail before we generalize it to arbitrary quantum algorithms. For the consideration of the time evolution we need to make a slight restriction to the RAGE states in that we do not consider general single qubit rotations but only those that may be absorbed in the MPS or WGS part of the state. Thus, we restrict attention to state vectors of the form
\begin{multline}
\ket{\psi(X,\varphi)} = \sum_{i_1,\ldots, i_N} \Tr{ X_{i_1}^{(1)}\ldots X_{i_N}^{(N)} }\\
\times \prod_{k,l} e^{i\delta_{i_k,1}\delta_{i_l,1}\varphi_{k,l}} \ket{i_1,\ldots, i_N}
\end{multline}
where we sum again over indices that are occurring twice. The action of a controlled-phase gates will affect only the adjacency matrix, \ie the values $\varphi_{k,l}$. Thus, only the action of the single qubit gates will require, perhaps surprisingly, any special attention. If all the phases were zero then the action of a single qubit unitary would be trivial as well as they would translate into a simple transformation of the type
\be
X_{i_k}^{(k)}\mapsto \sum_{j_k} X_{j_k}^{(k)}U_{i_k,j_k}.
\ee
However, the combination of the MPS with the WGS picture requires some extra thought. 

Given some initial state vector $\ket{\psi(Y,\varphi)}$, we assume that a single qubit gate $U_1$ acts on the first qubit. 
Then we are looking for a RAGE state vector $\ket{\psi(A,\varphi+\Delta\varphi)}$ which possesses the largest overlap with $U_1\ket{\psi(Y,\varphi)}$, \ie we would like to maximize
\be
\frac{|\bra{\psi(A,\varphi+\Delta\varphi)} H_1 \ket{\psi(Y,\varphi)}|^2}
{\braket{\psi(A,\varphi+\Delta\varphi)}{\psi(A,\varphi+\Delta\varphi)} \braket{\psi(Y,\varphi)}{\psi(Y,\varphi)}}.
\label{inner}
\ee
One may try and ignore the need for an update of the adjacency matrix, \ie set $\Delta\varphi=0$. One would however expect to obtain an approximation of better quality when also updating the adjacency matrix. In the following we keep the formulation as general as possible. To this end we will need to work out how to compute inner products between MGS-WGS vectors with differing adjacency matrices. This is only possible under certain constraints, namely restricting the variation to at most a single row of the adjacency matrix.

\subsubsection{Update of MPS-matrices} 

Let us assume for simplicity that the single qubit unitary $U_1$ is acting on the first qubit. Then we can disregard all entries in the adjacency matrix that do not affect qubit 1. For later purposes it will be most helpful to note that
\begin{multline}
\ket{\psi(Y,\varphi)}=
\sum_{i_2,\ldots,i_N} \Tr{Y_0^{(1)}Y_{i_2}^{(2)}\ldots Y_{i_N}^{(N)}}
\ket{0,i_2,\ldots ,i_N} \label{decomp}\\
+ \one \otimes \prod_{k=2}^N R_N(\varphi_{1,N})
\sum_{i_2,\ldots,i_N} \Tr{Y_1^{(1)}Y_{i_2}^{(2)}\ldots Y_{i_N}^{(N)}}
\ket{1,i_2,\ldots, i_N}
\end{multline}
where $R(\varphi)= \mathrm{diag}[1,\exp(i\varphi)]$ is a single qubit phase gate. Thus the action of the controlled phase gates can always be transformed into a set of single qubit operations which are then easily incorporated into the matrix product picture as
\begin{multline}
\ket{\psi(Y,\varphi)} = \sum_{i_2,\ldots,i_N} \Tr{Y_0^{(1)}Y_{i_2}^{(2)}\ldots Y_{i_N}^{(N)}} \ket{0, i_2,\ldots ,i_N} \\
+\sum_{i_2,\ldots,i_N} \Tr{Y_1^{(1)}Y_{i_2}^{(2)}(\varphi_{1,2})\ldots Y_{i_N}^{(N)}(\varphi_{1,N})} \ket{1, i_2,\ldots, i_N}
\end{multline}
This comes at the expense of a two-fold overhead in computational cost. This overhead increases exponentially with the number of qubits the original unitary $U$ is acting upon. Then we find
\begin{widetext}
\bee
\begin{split}
\bra{\psi(A,\varphi+\Delta\varphi)}U_1\ket{\psi(Y,\varphi)} =& \bra{0}U\ket{0} \Tr{(A_0^{(1)})^*\otimes Y_0^{(1)} \prod_{k=2}^N \sum_{i_k} (A_{i_k}^{(k)})^*\otimes Y_{i_k}^{(k)}} \\
&+ \bra{0}U \ket{1} \Tr{(A_0^{(1)})^*\otimes Y_1^{(1)} \prod_{k=2}^N \sum_{i_k} (A_{i_k}^{(k)})^*\otimes Y_{i_k}^{(k)}(\varphi_{1,k})} \\
&+ \bra{1} U \ket{0} \Tr{(A_1^{(1)})^*\otimes Y_0^{(1)} \prod_{k=2}^N \sum_{i_k} (A_{i_k}^{(k)}(\varphi_{1,k}+\Delta\varphi_{1,k}))^*\otimes Y_{i_k}^{(k)}} \\
&+\bra{1} U \ket{1} \Tr{(A_1^{(1)})^*\otimes Y_1^{(1)} \prod_{k=2}^N \sum_{i_k} (A_{i_k}^{(k)}(\varphi_{1,k}\Delta\varphi_{1,k}))^*\otimes Y_{i_k}^{(k)}(\varphi_{1,k})}    
\end{split}
\eee
\end{widetext}

Now we proceed in a two step procedure. First one picks an MPS-matrix for a single site $k$ and for fixed $\Delta\varphi$. This implies that the right hand side of the above expression is purely linear in the entries of the matrices $A^{(k)}$ belonging to site $k$, \ie no constant terms appear. Then, we are in a position to translate this maximization of the overlap into a generalized eigenvalue problem, \ie
\be
\frac{|\bra{\psi(A,\varphi+\Delta\varphi)} H_1 \ket{\psi(Y,\varphi)}|^2}
{\braket{\psi(A,\varphi+\Delta\varphi)}{\psi(A,\varphi+\Delta\varphi)}\braket{ \psi(Y,\varphi)}{\psi(Y,\varphi)}}
\ee
can now be formulated as a purely quadratic form in the numerator and denominator and thus be solved by a generalized eigenvalue problem.
The other free variables are the angles $\Delta\varphi_{1,l}$ for $l=1,\ldots,N$ which also need to be updated. How this is done will be explained in the following.

\subsubsection{Update of the adjacency matrix} 

One simple, though not very efficient, approach would be to randomly pick a $\Delta\varphi_{1,l}$ and vary its value accepting it when one found an improvement. Such an approach is however slow and prone to local minima. Thus we follow a slightly different approach that allows us to formulate the problem again as a generalized eigenvalue problem. Let us update the entry of the adjacency matrix between qubit $1$ and $k$ and assume that $\Delta\varphi$ has non-zero entries only in the first row and column. Now consider the non-unitary operator 
\be
U_{1,k} &=& \text{diag}[a+ib,a-ib,a-ib,a+ib] \\
&=& a\one+ib\sigma_z\otimes\sigma_z.
\ee
Then, with the same unitary $U_1$ as before we find
\begin{equation}
\frac{|\bra{\psi(A,\varphi+\Delta\varphi)} U^{\dagger}_{1,k} U_1 \ket{\psi(Y,\varphi)}|^2} {\bra{\psi(A,\varphi+\Delta\varphi)} U^{\dagger}_{1,k} U_{1,k} \ket{\psi(A,\varphi+\Delta\varphi)} \bra{\psi(Y,\varphi)}{\psi(Y,\varphi)}}
\label{opt}
\end{equation}
All terms are easy to evaluate and we will only discuss a non-trivial term arising in the evaluation of the numerator here. We find
\bee
\begin{split}
&\bra{\psi(A,\varphi+\Delta\varphi)}U^{\dagger}_{1,k} \sigma_x^{(1)} \ket{\psi(Y,\varphi)} \\
&= \bra{\psi(A,\Delta\varphi)} U^{\dagger}_{1,k} \sigma_x\otimes R_{\varphi_{1,2}} \otimes \ldots \otimes R_{\varphi_{1,N}}\ket{\psi(Y)} \\
&= \bra{\psi(A,\Delta\varphi)} U^{\dagger}_{1,k} \ket{\psi(Y_x)}\\
&= a \braket{\psi(A,\Delta\varphi)}{\psi(Y_x)} \\
&\quad + i b \bra{\psi(A,\Delta\varphi)} \sigma_{z}^{(1)} \otimes \sigma_z^{(k)} \ket{\psi(Y_x)}
\end{split}
\eee
Thus we obtain an expression that is purely linear in $a$ and $b$ leading to a generalized eigenvalue problem which can thus solve it efficiently. The evaluation of individual terms proceeds in analogy to the decomposition used in Eq.~\eqref{decomp}.

\subsubsection{Further improvements} 

It is clear from the above that we have to restrict the update to a single row (and the corresponding column) in the adjacency matrix. The computational effort scales exponentially with the number of rows that are updated in one step. This may still seem like an undue simplification that may restrict the success of the method. Let us consider the following improvement. Take a single qubit gate $U=e^{iH}$ where $H$ is a Hermitian matrix. So far we have optimized it in one step. Instead let us now use that $U=(e^{iH/N})^N$ and adopt the following strategy. Apply $e^{iH/N}$ to the first qubit and then update the $1$-st column. Now apply $e^{iH/N}$ to the first qubit and then update the 2nd column. In the $k$-th step apply $e^{iH/N}$ to the first qubit and then update the $k$-th column. Continue until you reach $k=N$.

\subsubsection{Hamiltonian time-evolution}

The above simulation of quantum algorithms happens in discrete time applying discrete quantum gates. 
A very similar approach may be taken for Hamiltonian time evolution whereby one uses a {\it Trotter expansion} to
represent the matrix exponential of $H$ \cite{DKSV04,S05}. As $H$ is a sum of operators with bounded support we are now able to compute inner products of the form Eq.~\eqref{inner} and thus simulate the time evolution. Now expanding the individual terms in the Hamiltonian $H$ and employing the above relation to commute terms we can simplify the inner products in a manner analogous to the previous section and thus obtain simple generalized eigenvalue problems that allow us to update efficiently both the matrices in the matrix product state and the entries of the adjacency matrix.  

\subsubsection{Promising simple quantum circuits}

Apart from full scale quantum algorithms such as Shor's factoring algorithm
there are some simple applications that require only a moderate set of gates.
Let us begin with the discussion of a particularly simple quantum algorithm, the {\it quantum Fourier transform}, 
which highlights the potential of the combined MPS-WGS description. Now, it is known that several instances of the quantum Fourier transform are classically efficiently simulatable:
This applies to the semi-classical quantum Fourier transform in which the outcome of the circuit is measured in the computational basis 
\cite{GN96}. Also, the approximate quantum Fourier transform -- in which phase gates with small
phases are neglected -- can be efficiently simulated by exploiting a tree tensor network \cite{YS07}.
Ref.\ \cite{ALM06} considers the simulation of the exact
quantum Fourier transform, showing that  
if a quantum state can be generated with a small bubble width circuit, and if the
Fourier transform subroutine does not increase the so-called bubble width significantly, 
then the Fourier coefficients of the state can be calculated efficiently classically (compare also Ref.\ \cite{N09}).

\begin{figure}[tb]
\centerline{\includegraphics[width=0.8\columnwidth]{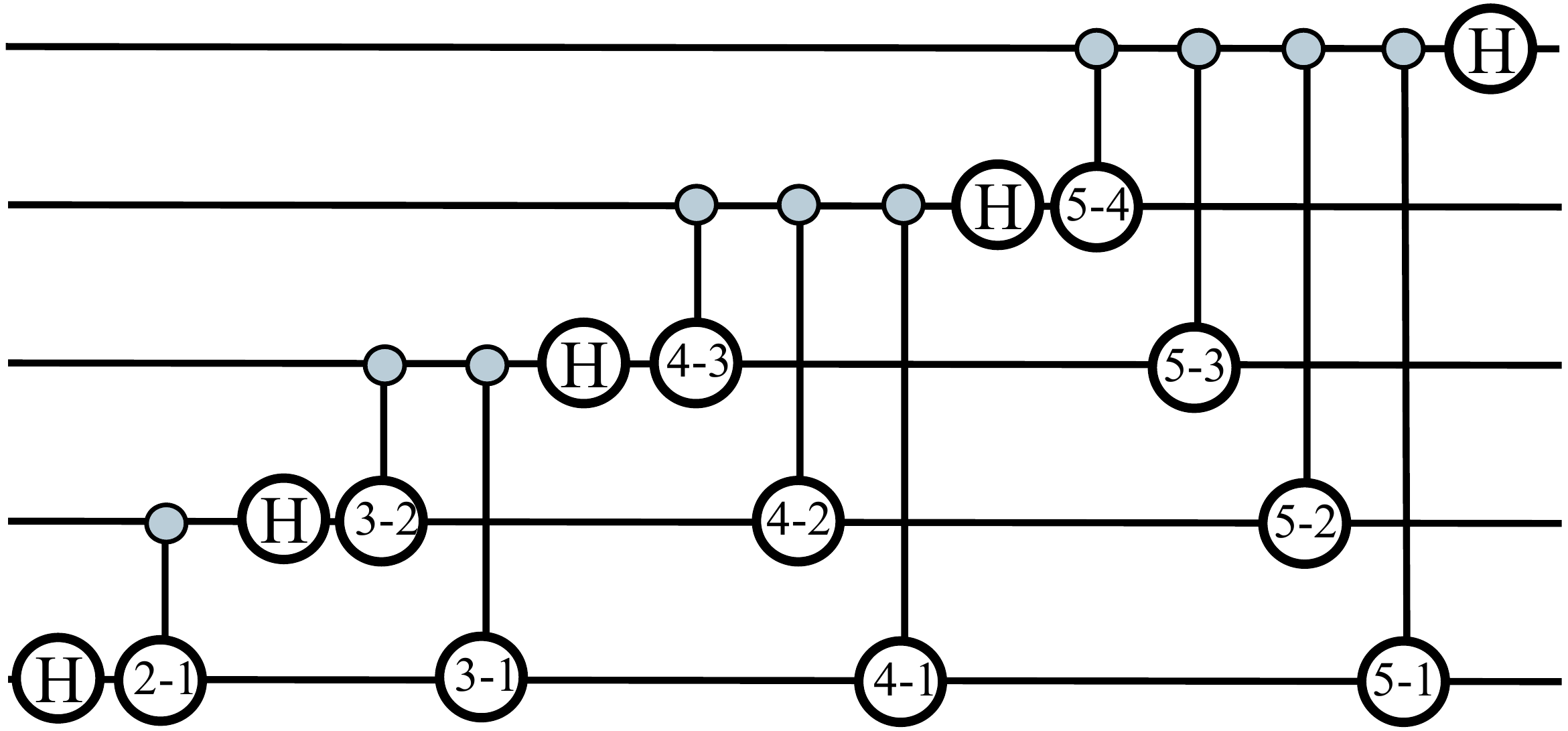}}
\caption{\label{DFT}The discrete Fourier transform is implemented by a sequence of Hadamard gates and controlled-phase gates with a rotation angle for a gate between qubits $i$ and $j$ of $\pi/2^{i-j}$. Time increases from left to right.}
\end{figure}

Still the quantum Fourier transform
constitutes a valid setting in which a RAGE approach is expected to perform very favorably compared to an approach based
on MPS. The gate sequence realizing the discrete Fourier transform is given in Fig.\ \ref{DFT}.
Only Hadamard and controlled-phase gates are being used. This already suggests that MPS-WGS states are ideally suited for the simulation of this algorithm. 
Note that an MPS approach alone is expected to deliver not very good results: although the angles in the controlled phase gates are in parts becoming very small for large $N$, the state at intermediate times does not satisfy asymptotically an area law in $N$ in bisections. This has been looked at numerically for moderate values of $N$. 

Another more feasible and very exciting application of RAGE would be the phase estimation problem. In a way, any known quantum algorithm of the hidden subgroup variant is essentially a phase estimation problem. The task is, given an $n$-qubit unitary $U$, and a single eigenvector $\ket{v}$, estimate the phase of the eigenvalue
\be
U \ket{v} = e^{-2\pi i \phi} \ket{v}.
\ee
In the actual quantum version, we are thought not to know $U$ or $|v\rangle$ or $e^{-2\pi i \phi}$, but one is assumed to have devices that perform a controlled-$U$, and $U^{2^1}$, and $U^{2^2}$, and so on.
To do so, prepare first 
\be
\ket{\psi_0} =\ket{0,\ldots,0}\ket{v},
\ee
where the first $m$ qubits are prepared in the $\ket{0}$ state, so $m+n=N$. Then, apply Hadamards to this state vector, 
\be
(H^{\otimes m} \otimes \one ) \ket{\psi_0} = \ket{0,\ldots,0}\ket{v}.
\ee
How to do this will be explained below. This step is followed by the controlled unitaries $\ketbrad{0} \otimes \one + \ketbrad{1} \otimes U^{2^k}$. We here should assume that we can conveniently decompose these controlled unitaries in a network of matrices that can be meaningfully decomposed in our setting. This is then followed by an inverse DFT. This could be a setting of a full quantum algorithm where we can estimate local properties finally (or, for that matter, the phase).

\subsection{RAGE states based on TTS}\label{sec:TTSandWGS}

In this subsection, we will extend our prior results and present and discuss the RAGE states based on TTS as well as applicable algorithms. In the case of closed-boundary MPS, we exploited the matrix product structure underlying the MPS tensor-network in order to make the algorithms efficient. Similar to the original formula for the computation of the entries in the reduced density matrix in an MPS, the corresponding formula for the RAGE state is based on products of transfer matrices, with a modification of phases that are derived from the WGS part. Analogously, also RAGE states based on TTS inherit the algorithms from the TTS, with some modifications. This section contains two new aspects of the graph enhancement, as compared to the implementation given for MPS. First, we will generalize the application of phase gates from qubits to higher dimensional qudits. Second, we will (in the context of TTS) create intermediate auxiliary observables in order to implement the graph enhancement and to compute the entries of the reduced density matrix. Hence, the effect of the WGS operators is not absorbed by the tensor-network state anymore, as in the case of the MPS implementation given before.

Let us first define the generalized phase gate, acting on two $q$-level spin systems named $a$ and $b$
\be
W(\varphi) = \sum_{s_1,s_2 = 0}^{q-1} \ketbrad{s_1 ,s_2} e^{i \varphi[s_1,s_2]}.
\label{eq:Wdef}
\ee
Here, $\varphi$ is a symmetric matrix $\varphi = \varphi^T$, and $\varphi[0,s]=0$ for all values of $s$. For the whole $N$-particle system to be described we need one of these matrices for each pair of sites, hence $\varphi$ is a tensor of rank four with elements denoted by $\varphi_{a,b}[s_1,s_2]$. With it we define the operator
\be
\mathcal{W}_{\varphi}=\prod_{a<b} W^{(a,b)}(\varphi_{a,b}).
\label{eq:WWdef}
\ee
With this operator and local rotations $V_j\in U(2)$ we define a RAGE state corresponding to a tree tensor state $\ket{\psi_{\text{TTS}}}$ with tree $\tau$ over $N$ sites as
\be \label{eq:defRAGETTS}
\ket{\psi(\tau,\varphi,V)}:=\mathcal{W}_{\varphi}\ket{\psi_{\text{TTS}}}
\ee
These states inherit the properties of the RAGE states based on MPS, as the MPS with open boundary conditions are a special case of the TTS. These states offer a more variable geometry of the graph corresponding to the tensor product state, thus leading to RAGE states better suited for certain systems, see Sec.~\ref{sec:comparisonMPSTTS}.

\subsubsection{Efficient computation of local properties and correlation functions}\label{sec:TTSWGScomp}

We will now derive an efficiently computable expression for the reduced density matrix of the RAGE state $\ket{\psi(\tau,\varphi,V)}$ corresponding to a TTS. This allows for an efficient computation of expectation values of small support, such as local Hamiltonians that are sums of polynomially many terms with small support, like \eg in Eq.~\eqref{eq:general hamiltonian}, and moreover correlators of small support.

We will consider the expression for a support of two sites $\{a,b\}=\mathcal{S}$ with complement $\bar{\mathcal{S}}$. The expression to be calculated is hence
\be
\rho_{\mathcal{S}} = \pTr{\bar{\mathcal{S}}}{\mathcal{W}_{\varphi} \ketbrad{\psi_{\text{TTS}}} \mathcal{W}_{\varphi}^{\dagger}}.
\ee
The cyclicity of the trace causes all phase operators with support in $\bar{\mathcal{S}}$ to cancel. Hence we can rewrite the expression above using the operator
\be
\mathcal{W}_{a,b} := W^{(a,b)}(\varphi_{a,b}) \prod_{c \in \bar{\mathcal{S}}} W^{(a,c)}(\varphi_{a,c}) W^{(c,b)}(\varphi_{c,b})
\label{eq:WWabdef}
\ee
instead, and find that the matrix elements of $\rho_{\mathcal{S}}$ are given by
\begin{multline}
\bra{s'_a, s'_b} \rho_{\mathcal{S}} \ket{s_a ,s_b} = \\
\bra{\psi_{\text{TTS}}}\left[ \mathcal{W}_{a,b} (\ketbra{s_a}{s'_a})^{(a)} (\ketbra{s_b}{s'_b})^{(b)} \mathcal{W}_{a,b}^{\dagger} \right]\ket{\psi_{\text{TTS}}}.
\end{multline}
So, each matrix element of the reduced density matrix is given by the expectation value of an observable evaluated in a TTS. As it is a product of local observables, we can use methods known for TTS, see also Sec.~\ref{sec:TTS}, to compute it efficiently.

To see this we define now, using a vector $\phi$, the local operator
\be
V(\phi) := \sum_{s=0}^{D-1} \ketbrad{s} e^{i \phi_s}
\ee
and notice that the operators $W^{(a,b)}(\varphi_{a,b})$ can be written as
\be
W^{(a,b)}(\varphi_{a,b}) = \sum_{s=0}^{d-1} \ketbrad{s} \otimes V(\varphi_{a,b}[\cdot,s]).
\ee
In this expression, $\varphi_{a,b}[\cdot,s]$ denotes the $s$-th column/row of the symmetric matrix $\varphi_{a,b}$. Using this 
expression, we have
\begin{multline}
\mathcal{W}_{a,b}=\ketbrad{s_a}^{(a)} \ketbrad{s_b}^{(b)} e^{i \varphi_{a,b}[s_a,s_b]}\\
\times \prod_{c \in \bar{\mathcal{S}}} \biggl[ \left(\sum_{s_a=0}^{D-1}(\ketbrad{s_a})^{(a)} V(\varphi_{a,c}[\cdot,s_a])^{(c)}\right)\\
\times \left(\sum_{s_b=0}^{D-1}(\ketbrad{s_b})^{(b)} V(\varphi_{b,c}[\cdot,s_b])^{(c)}\right)\biggr],
\end{multline}
and using the relations $V(\phi_1) V(\phi_2) = V(\phi_1+\phi_2)$ and $V(\phi)^{\dagger} = V(-\phi)$, we arrive at
\begin{widetext}
\begin{multline}
\bra{s'_a ,s'_b} \rho_{\mathcal{S}} \ket{s_a ,s_b} \\
= \bra{\psi_{\text{TTS}}} \left[(\ketbra{s_a}{s'_a})^{(a)} (\ketbra{s_b}{s'_b})^{(b)} \prod_{c \in \bar{\mathcal{S}}} V^{(c)}(\varphi_{a,c}[\cdot,s_a] -\varphi_{a,c}[\cdot,s'_a] +\varphi_{b,c}[\cdot,s_b] -\varphi_{b,c}[\cdot,s'_b]) \right] \ket{\psi_{\text{TTS}}} \\
\times \exp{i \left[ \varphi_{a,b}[s_a,s_b] -\varphi_{a,b}[s'_a,s'_b]\right]}. \label{eq:RAGEevalTTS}
\end{multline}
\end{widetext}

This expression can be calculated using the known methods from Sec.~\ref{sec:observable evaluation in TTN} with an effort of $O(N D^6)$ for each element of $\rho_{\mathcal{S}}$. Hence, as in the case of MPS, the augmentation with WGS did not increase the computational effort when evaluating the reduced density matrix.

\subsubsection{Variational methods}\label{sec:TTSandWGSupdates}

The optimization of expectations values in RAGE states based on TTS is very similar to the procedures in RAGE states based on MPS. We optimize the tensorial part and the phases independently in a sweeping procedure.
Because the relation
\be
\bra{\psi(\tau,\varphi,V)}O^{(\mathcal{S})}\ket{\psi(\tau,\varphi,V)}
= \Tr{O \rho_{\mathcal{S}}}.
\ee
holds, and all matrix elements of $\rho_{\mathcal{S}}$ use the same underlying tensor product state $\ket{\psi_{\text{TTS}}}$, the computation of an observable in a RAGE vector $\ket{\psi(\tau,\varphi,V)}$ amounts to the computation of an expectation value of a linear combination of observables of the kind as given above in Eq.~\eqref{eq:RAGEevalTTS}, in the tensor product vector $\ket{\psi_{\text{TTS}}}$. Provided that the support of the operator $O^{(\mathcal{S})}$ is small, the number of terms in the linear combination of operators applied to $\ket{\psi_{\text{TTS}}}$ remains small as well.

For the tensorial part, similar to the procedure applied to bare TTS, we consider the state vector $\ket{\psi_{\text{TTS}}}$ as a linear combination of other RAGE states, where the combination is controlled by one of its tensors, say $A$, see Eq.~\eqref{eq:TTS with T singled out}. The key step is then to compute the effective operators $\tilde H$ and $\tilde \one$ in the subspace spanned by the linear combinations, dependent on the tensor elements $A_{\alpha ,\beta, \gamma}$, and to solve the implied generalized eigenvalue problem for the Rayleigh quotient as for bare TTS.

The optimization of the phases can be realized by following the ideas applied in the case of MPS. In an initial step, we interpret the expectation value as a function of the parameters of the local unitaries and of the phases, and optimize with Nelder-Mead or gradient based methods. Following this initial step, we can use local optimizations very similar to the ones applied to MPS in a sweeping procedure. The generalization of the local phase gate $\Lambda Z^{(a,b)} (\varphi_{a,b})$ (with a single real variable $\varphi_{a,b}$) to operators $W(\varphi)$ (with real symmetric matrices $\varphi$), acting on $d$-level systems, implies a similar relation as in Eq.~\eqref{eq:operator sincosrelation}. While the operator is still linear in all sine-- and cosine functions, the main difference is that more than one phase appears in the dissection of the operator $W(\varphi)^{\dagger} \tilde H_{c,d}^{(c,d)} W(\varphi)$ corresponding to $\Lambda Z^{(a,b)} (-\varphi_{a,b}) \tilde H_{c,d}^{(c,d)} \Lambda Z^{(a,b)} (\varphi_{a,b})$ in Eq.~\eqref{eq:operatortodissect}. All the phases can still be optimized one-by-one, relying an expression analogous to Eq.~\eqref{eq:singlephase endform}. The comments on the unsatisfactory optimization speed of the phase optimization in the context of MPS applies also in this case.

\subsection{Further extensions}\label{sec:extension}

In this section we will focus on possible extensions of the RAGE method. 
Consider a class of state vectors of the form
\be
\label{gRAGE}
\ket{\psi} = {\cal U} \ket{\varphi},
\ee
where $\ket{\varphi}$ stands for a quantum state vector described by some tensor-network for which an (approximate) algorithm to evaluate tensor products of local observables $O_1 \otimes O_2 \ldots O_N$ exists. For instance, MPS, TTS and PEPS fall within this class. In addition, ${\cal U}$ denotes a circuit of quantum gates to be specified below. In order to find the optimal representative within such a variational class of states that minimizes the energy, $\min_{\ket{\psi}}\bra{\psi}H\ket{\psi} / \braket{\psi}{\psi}$, in many relevant systems it suffices to compute expectation values of observables with a small support. In particular, for pairwise interaction Hamiltonians the support is limited to two, as the Hamiltonian can be written as a sum of (at most polynomially many) terms of the form $O_{1}^{(\alpha)}O_{2}^{(\beta)}$ acting only on particles $\alpha$ and $\beta$. In fact, any such observable can be written in the Pauli basis, so it suffices to restrict oneself to Pauli operators $\sigma_i^{(\alpha)}\sigma_j^{(\beta)}$ as $O_{1}^{(\alpha)}O_{2}^{(\beta)}$ can be represented by a sum of at most 16 Pauli-terms. Similar considerations hold for any $k$-body interaction Hamiltonian with bounded $k$.

Consider now $\bra{\psi}\sigma_i^{(\alpha)}\sigma_j^{(\beta)}\ket{\psi}$ with $\ket{\psi}$ being of the form Eq.\ \eqref{gRAGE}. We let ${\cal U}$ act on the observable and obtain $\bra{\varphi} ({\cal U^\dagger} \sigma_i^{(\alpha)} \sigma_j^{(\beta)} {\cal U}) \ket{\varphi}$. It is immediately clear that as long as ${\cal U}$ transforms the two Pauli operators to either a tensor product of Pauli operators or to a sum of polynomially many Pauli operators, this quantity can still be evaluated efficiently. Each of the terms only requires the evaluation of a tensor product of local observables for a tensor-network state, which is efficiently possible by assumption. Such a situation occurs for example for {\em any} operator ${\cal U}$ corresponding to a {\em Clifford circuit}, which maps by definition tensor products of Pauli operators to tensor product of Pauli operators under conjugation.

Notice that a Clifford circuit can thereby produce a large amount of entanglement and is not restricted in its depth. Moreover, a Clifford circuit is able to enlarge the support of an operator arbitrarily; hence an evaluation of an observable under such a modification is in general not efficient using the reduced density matrix. Algorithms allowing for the evaluation of product observables with arbitrary support, as given for TTS before, are still applicable.

Similarly, any quantum circuit of depth $n$ consisting of nearest-neighbor gates transforms the initial observable with support two to an observable that acts (at most) on $4n+2$ particles and hence can be represented by a sum of at most $4^{4n+2}$ Pauli terms. As long as the depth $n$ of the circuit is bounded or scales logarithmically in the number of particles $N$, one can efficiently evaluate the resulting observable as a sum of only polynomially many Pauli-terms. Hence all states of the form Eq.\ (\ref{gRAGE}) where ${\cal U}$ is a log-depth quantum circuit and $\ket{\varphi}$ is \eg a MPS, TTS or PEPS can be used as variational family.

Notice that also the efficient observable evaluation for the RAGE class based on MPS discussed in the previous section can be understood in this terms, although we have given a more efficient implementation tailored for operators with small support there. In this case, 
\begin{equation}
	{\cal U}=\prod_{\gamma,\delta} U_{\rm PG}^{(\gamma,\delta)} 
\end{equation}
is a product of commuting phase gates. Hence all phase gates that do not act on particles $\alpha$ or $\beta$ cancel in ${\cal U^\dagger} \sigma_i^{(\alpha)}\sigma_j^{(\beta)} {\cal U}$. The action of the remaining gates acting on particles $\alpha,\beta$ can however be described by a matrix product operator (or other tensor-network operator if required) of small dimension. This can be easily seen by considering a bi-partition of the system into particles $\alpha,\beta$ vs. rest. All gates that do not cancel act between these two groups (or between $\alpha$ and $\beta$). The amount of entanglement between these two groups is bounded by the physical dimension $d^2$ of the system $\alpha, \beta$, and hence also the amount of entanglement of any operator acting between these two subsystems is bounded by $d^4$. For phase gates one finds that the resulting operator describing ${\cal U^\dagger} \sigma_i^{(\alpha)}\sigma_j^{(\beta)} {\cal U}$ can be written as a matrix product operator of dimension $D=4$. Applying such an operator to a matrix product state of dimension $D'$ yields another matrix product state of dimension at most $4D'$. The evaluation of the observable $\sigma_i^{(\alpha)}\sigma_j^{(\beta)}$ hence reduces to calculating the overlap between two matrix product states of (slightly) increased dimension, which still can be done efficiently.

So far, RAGE states have been used to implement an enhancement of tensor network states due to the incorporation of features of weighted graph states. Now we would like to consider the {\it entire class of Clifford circuits}. Thereby, we will perform the whole step of shifting the effect of the enhancement away from the picture of the modification of the underlying tensor-network state to the new picture where the observables themselves are modified by the enhancement. Please note that extensions into the other direction, namely the extension to more general tensor-network states, are also possible, as will be demonstrated at the example of PEPS states in Appendix~\ref{sec:PEPS}.

As we have seen in the previous sections, several families of tensor-network states can be supplemented with circuits of commuting unitary operations in such a way that they form a good variational set of states. In particular, relevant quantities such as expectation values of local observables (\eg energy of $k$-body interaction Hamiltonians) can be efficiently computed, and efficient optimization methods are applicable. In this section we show that one can indeed replace the circuits of commuting phase gates by other types of quantum circuits such as, \eg Clifford circuits while maintaining the favorable features regarding evaluation and optimization, thereby significantly extending the variational class of states, their respective features and their possible range of applicability.

\section{Applications of RAGE states}\label{sec:appofRAGE}

In this section we want to consider applications of RAGE states based on different tensor product states as well as bare tensor product states in comparison. There is a wide range of possible applications for variational quantum states. Among them is the numerical exploration of quantum mechanical condensed matter systems. As theoretical considerations and real world experience suggest, ground states of interacting quantum systems typically show emergent phenomena, often due to a high degree of correlations of the local constituents. It is hence challenging to provide variational classes that are able to grasp these correlations and entanglement features. 

To investigate the ability of the RAGE states to approximate ground states, we treat the Ising and Heisenberg model in two dimensions. When describing the ground states of these Hamiltonians, the MPS and TTS usually perform badly in the regime where the dimensions of the employed matrices allow for an efficient treatment. This behavior is expected to be improved by introducing the WGS enhancement to the MPS and TTS, allowing essentially to follow an area-law for the entanglement. In the performed simulations, the adjacency matrix is allowed to connect any constituents in the lattice with individual phases. This is the most general enhancement that we can implement with our RAGE ansatz, however, it is possible to introduce more symmetries and restrictions to speed up the computation.

\subsection{The 2D Ising model}

A well studied toy-model is the 2D Ising model with transverse magnetic field
\be
H=J \sum_{\langle a,b\rangle } \sigma_z^{(a)}\sigma_z^{(b)} + B\sum_a\sigma_x^{(a)}.
\ee
The aim of our numerical treatment is to find an approximation of the ground state by optimizing over 
the energy of trial states in a sweeping procedure.
\begin{figure}[t]
\includegraphics[width=0.9\columnwidth]{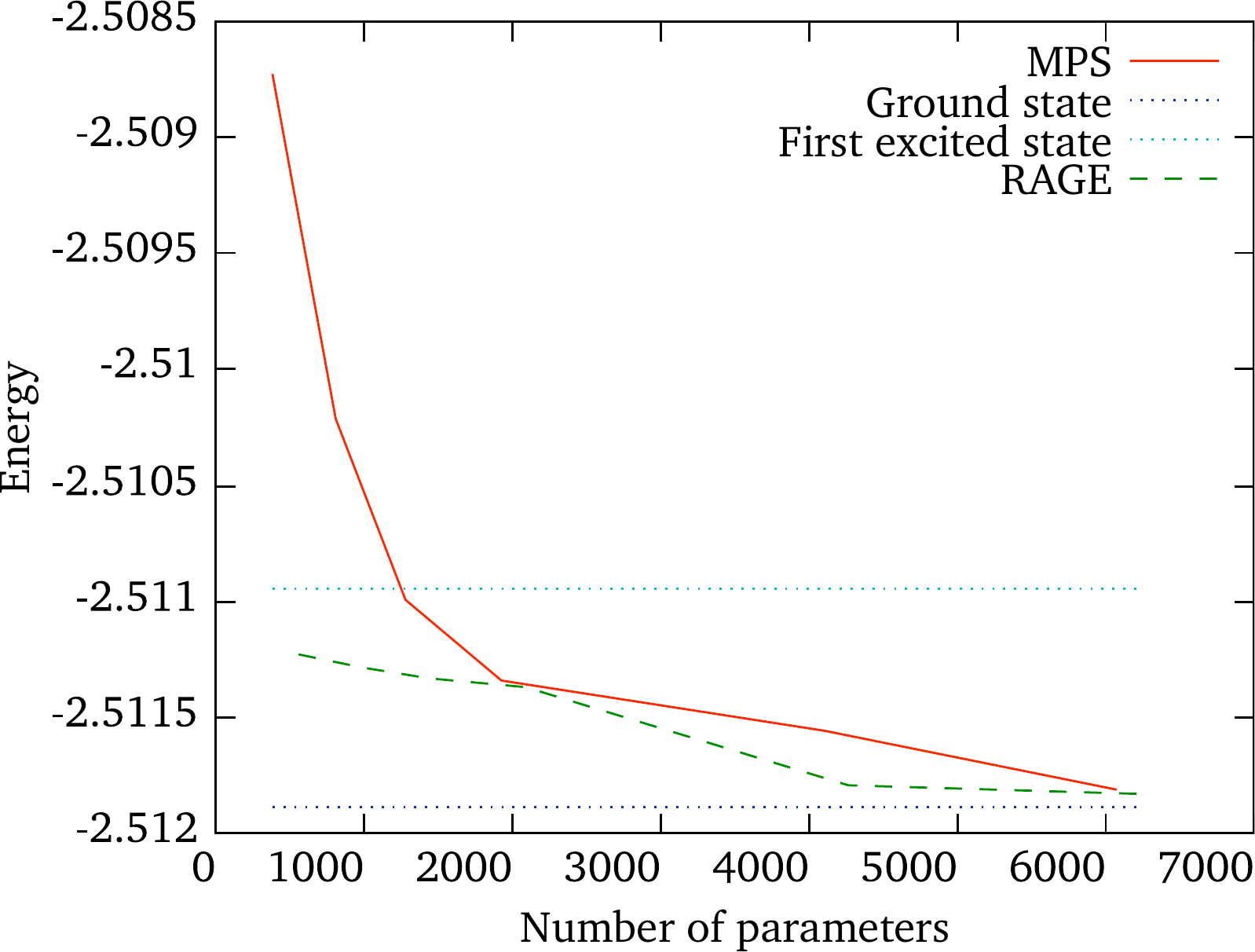}
\caption{(Color online) 2D Ising model with $B=2$ on a $4\times4$ periodic lattice. 
We compare the achieved accuracy within the RAGE approach and for MPS. 
The graph depicts the energy for different {\em total} numbers of complex parameters in comparison with the exact ground state as well as the first excited state.
In Figs.~\ref{fig:isingen}, \ref{fig:isingcor}, \ref{fig:heisenberg} the MPS is realized numerically as a \emph{flat} tensor tree state, \emph{i.e.} it is non-periodic. This implies that tensors close to the leaves of the tree change the local bases only and are hence redundant.}
\label{fig:isingen}
\end{figure}
Fig.~\ref{fig:isingen} shows the achieved ground state energy for the MPS alone and the RAGE states based on MPS, where in both cases the MPS follows a chain-like path through the 2D setting. What can be seen is that for a very low parameter count, the RAGE states perform better than the MPS. However, this advantage becomes negligible for larger numbers of parameters, although qualitatively the RAGE states always perform better. In the regime with a small number of parameters, \eg around $300-1000$, the actual quality of the approximation is not very good, as the states approximating the ground state have an energy comparable to the first excited state.

\begin{figure}[t]
\includegraphics[width=0.9\columnwidth]{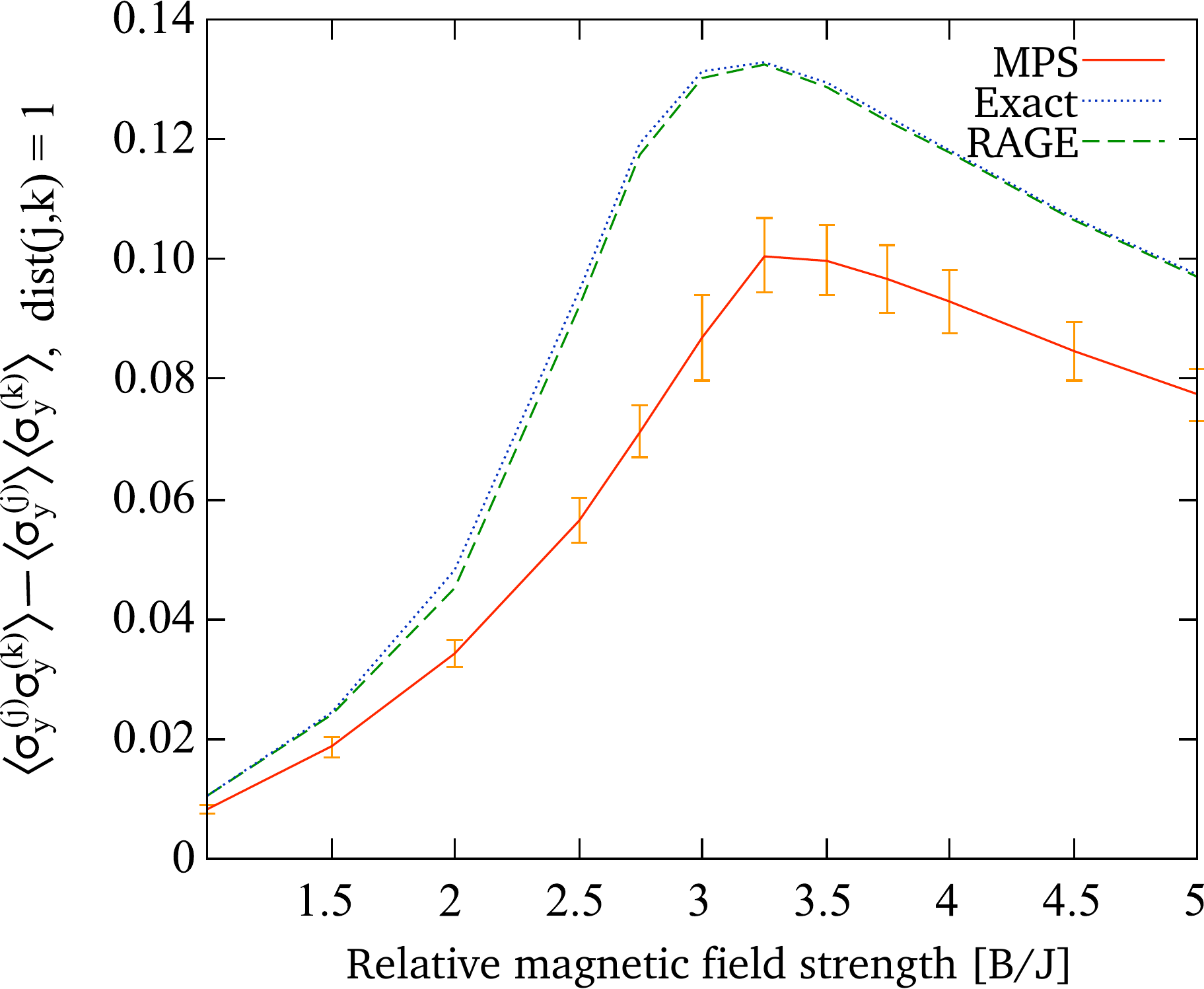}
\caption{(Color online) 2D Ising model on a $4\times4$ periodic lattice. We compare the achieved accuracy using RAGE and MPS ($D=4$) with exact results. The total number of independent parameters is $384$ for MPS and $476$ for the RAGE state. Two-point correlations as function of $B$ are shown.
The error bars stem from differences in the local two-point correlations due to the explicit breach of translational symmetry in the necessarily linear topology of the MPS ansatz.}
\label{fig:isingcor}
\end{figure}

The situation is very different when we look at the correlations. As Fig.~\ref{fig:isingcor} shows,
the RAGE states show significantly better two-point correlations, already for a very small number of parameters, delivering virtually indistinguishable numerical values compared to the exact results. The values shown in the graph are mean values over all sites (and their neighbors) in the system. The high variance of the values in the MPS case stems from the different values obtained at different sites, as the MPS description necessarily breaks the 2D symmetry of the model. The RAGE description does not show a any variance (within numeric accuracy), the correlations being translationally and rotationally invariant, as is the Hamiltonian. This shows that the RAGE states can adapt to the 2D setting much better and do not break the symmetry as the MPS alone.

It is interesting to see that the correlations are described well, even though the energy approximation is comparatively bad, although we aimed for a good description in terms of the energy. This seems to indicate that even in a situation where some features of the system cannot be approximated well by the RAGE states, geometry-dependent symmetries will be reflected. In numerics, the situation is often the opposite: Although a parameter that has been aimed at is well-described, other features of the system are completely lost in the description. The RAGE states seem to be favorable in a situation where geometric symmetries play a role.

\subsection{The 2D Heisenberg model}

Another toy model that we use is the 2D Heisenberg model, described by
\be
H=\sum_{\langle a,b\rangle } \sigma_x^{(a)}\sigma_x^{(b)} + \sigma_y^{(a)}\sigma_y^{(b)} + \sigma_z^{(a)}\sigma_z^{(b)},
\ee
where $\langle a,b\rangle$ denotes nearest neighbors. It is known to be numerically more challenging than the 2D Ising model. We approximate the ground state of the system using MPS and TTS, as well as RAGE states based on MPS and TTS.
\begin{figure}[t]
\includegraphics[width=0.9\columnwidth]{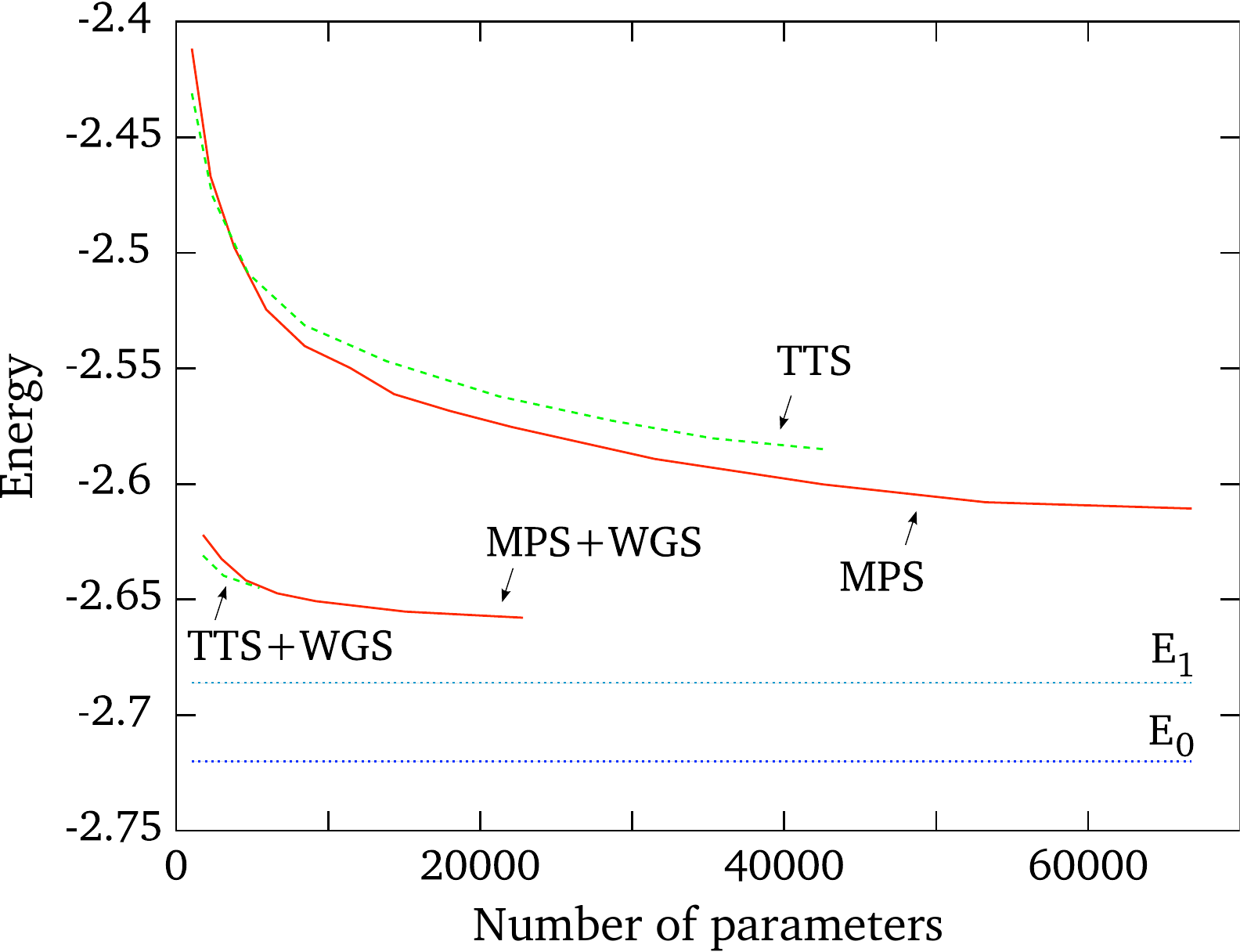}
\caption{(Color online) 2D Heisenberg model on a $6\times6$ periodic lattice. We compare the achieved accuracy of bare tensor-network states with RAGE states. Also, values of the ground state energy and the first excited energy from quantum Monte Carlo simulations are shown \cite{QMC}.
The number of computational steps for the evaluation of expectiation values is, as shown in the main text, $O(N D^5)$ and $O(N D^6)$ for MPS and TTS respectively; the combination of these states with WGS amounts to a change by a constant factor.
The optimization, in turn, relies on a constant number of evaluations of expectation values per single optimization step. The number of such steps needed for satisfactory convergence is much higher for RAGE states than bare MPS and TTS as the state space is different.}
\label{fig:heisenberg}
\end{figure}
We observe that the achieved energy is always larger than even the energy of the first excited state. However, as expected, the RAGE states perform better than the MPS and TTS without WGS enhancement, the TTS having a slight advantage in the regime with a low parameter count.

Even though the RAGE states allow for new features like long-range correlations and a violation of an area law, they break the local $SU(2)$ gauge invariance. It is clear from the simulations that the limitation of the underlying 1D structure of the MPS and TTS cannot always be fully overcome by the graph enhancement.

\subsection{Simulation of a disturbed toric code state}

The RAGE method is particularly well suited for certain interesting parent Hamiltonians, as the WGS include all graph and hence stabilizer states. Any graph state vector $\ket{G}$ is the common eigenvector with eigenvalue $1$ to a set of operators $\{ K_a \}$, \ie $K_a \ket{G} = \ket{G}$, the state vector $\ket{G}$ is automatically the ground state of the Hamiltonian $H_G:= - \sum_a K_a$. Imperfections, \eg stemming from imprecise preparatory procedures, can be simulated by a slight deviation from the given form, \eg with local magnetic fields
\be
H'_G:= - \sum_a K_a + \sum_i h_i \sigma_z^{(i)}.
\ee
The undisturbed forms of these states, \ie the ground states of the undisturbed Hamiltonians $H_G$, have phases that can be found analytically, using an algorithm relying essentially on Gaussian elimination. If the amount of noise is small, these phases are still reasonably close to the phases of the noisy states, \ie the ground states of the disturbed Hamiltonians $H'_G$. Hence, treating these systems, the values of the phases of the noisy states will be fixed to the analytically found values of the corresponding undisturbed state, and the approximation will be performed over the tensors only. This is useful insofar as the WGS operators contain the essential entanglement characteristics also of the slightly disturbed states, which leaves only the (disturbance of the) entanglement created by the noise terms to be reproduced by the tensor product state. In many noise models it is sufficient to use MPS or TTS with small index dimensions only, as the noise is typically spatially uncorrelated. A variational method based only on WGS does not offer this kind of ansatz, as all the entanglement has to be described by the WGS part. 

The example that we use is the perturbed Kitaev model, whose ground state is, in the unperturbed case, the toric code state~\cite{Ki03}. This state is actually a sub\emph{space} of the Hilbert space which is the common eigenstate with eigenvalue $1$ of a set of operators to be defined in the following. Starting from a rectangular lattice, we identify all edges between nearest-neighbor vertices in the lattice with quantum mechanical spins. A generator the operator set can now be defined by the operators $(\sigma_x \otimes \sigma_x \otimes \sigma_x \otimes \sigma_x)^{(\text{loop})}$ and $(\sigma_z \otimes \sigma_z \otimes \sigma_z \otimes \sigma_z)^{(\text{cross})}$ where the loops are the smallest possible loops (constituting the set $L$) in the lattice and the crosses the smallest possible edge configurations with a cross shape (constituting the set $C$) except for one, otherwise the generating set would not be independent. The unperturbed Hamiltonian whose ground state is given by all states in the stabilized subspace can hence be chosen as
\begin{multline}
H_{\text{Toric code}}= - \sum_{\ell \in L} (\sigma_x \otimes \sigma_x \otimes \sigma_x \otimes \sigma_x)^{(\ell)} \\
- \sum_{c \in C} (\sigma_z \otimes \sigma_z \otimes \sigma_z \otimes \sigma_z)^{(c)}.
\end{multline}
As the operators form an (incomplete, because a subspace is stabilized) stabilizer, we are able to derive the local-unitary equivalent graph form of the Hamiltonian to obtain
\bee
H_{\text{Toric code}} \approx - \sum_a K_a.
\eee

\begin{figure}[t]
\includegraphics[width=0.9\columnwidth]{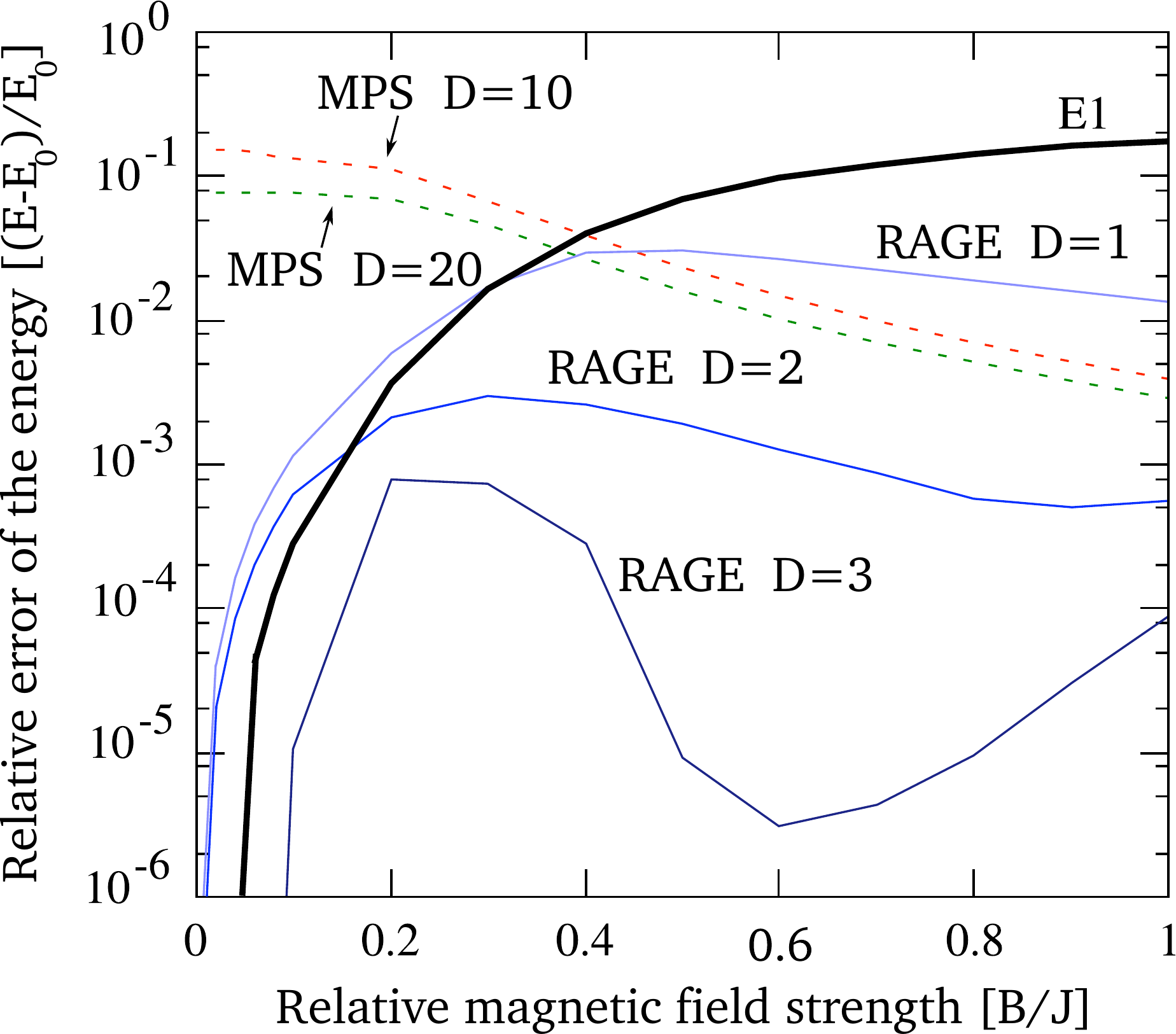}
\caption{(Color online) Kitaev model on a periodic 2D lattice with $N=12$ with magnetic field $B$  in x-direction. We compare the achieved accuracy for the ground state using MPS ($D=20$ and $D=10$) and RAGE states with fixed phases and underlying MPS of $D=1,2,3$, where $D=1$ corresponds to WGS. For $B=0$, the ground state is exactly described by a WGS. For comparison, the first excited state is plotted (E1). For larger $N$, similar results are found, although the exact treatment was no longer possible.}
\label{fig:toric}
\end{figure}

We have simulated the ground state of this model on a periodic 2D lattice with $12$ sites where the additional perturbation was described by uniform local magnetic fields of varying strength $B$, resulting in a Hamiltonian 
\begin{equation}
	H' = - J \sum_a K_a + B \sum_i \sigma_z^{(i)} 
\end{equation}	
describing the system under consideration. For numerical results, see Fig.~\ref{fig:toric}.
We observe that even for fixed phases of the WGS -- adjusted to match the toric code state at zero field -- and an underlying MPS with the small dimension $D=3$, we obtain a significantly improved accuracy as compared to WGS and MPS with a much higher number of parameters, \ie using a matrix dimension $D=20$. It is interesting to see that there is a local minimum for which the RAGE description is particularly good.

As the strength of the local magnetic field increases, the ground state of the Hamiltonian loses its similarity with the unperturbed ground state and eventually approaches a product state (although the field strength is not big enough to get really close to a product state.) Accordingly we observe that for small field strengths the RAGE states provide a good description of the ground state and the MPS cannot cope, even for high matrix dimensions. While the energy of the RAGE state with underlying matrix dimension $d=3$ is always well below the energy of the first excited state, the RAGE states with matrix dimension $d=1,2$ show this behavior only above a certain threshold. For strong local fields the MPS start to perform better and eventually outperform the RAGE state with $d=1$. This is due to the fact that the tensor-network part of the RAGE state has now to compensate for the wrong static phases (derived with assumption that local fields are absent) in the WGS description, which become worse estimates for the true phases as the influence of the local fields grows. However, the RAGE states with $d=2,3$ are always better than the bare MPS, even though the (now) wrong WGS phases must be compensated. We conclude that the RAGE states are better for the description of disturbed graph states than the bare WGS, which correspond to the case $D=1$.

This nurtures some hope that the RAGE description is actually powerful enough to cope with aversive situations, and that only the optimization algorithm that we employed for the ground state approximation could not find the optimal solutions. The improvement of these algorithms is therefore an important point for future research.

\subsection{Quantum circuits}

\begin{figure}[tb]
\includegraphics[width=.9\columnwidth]{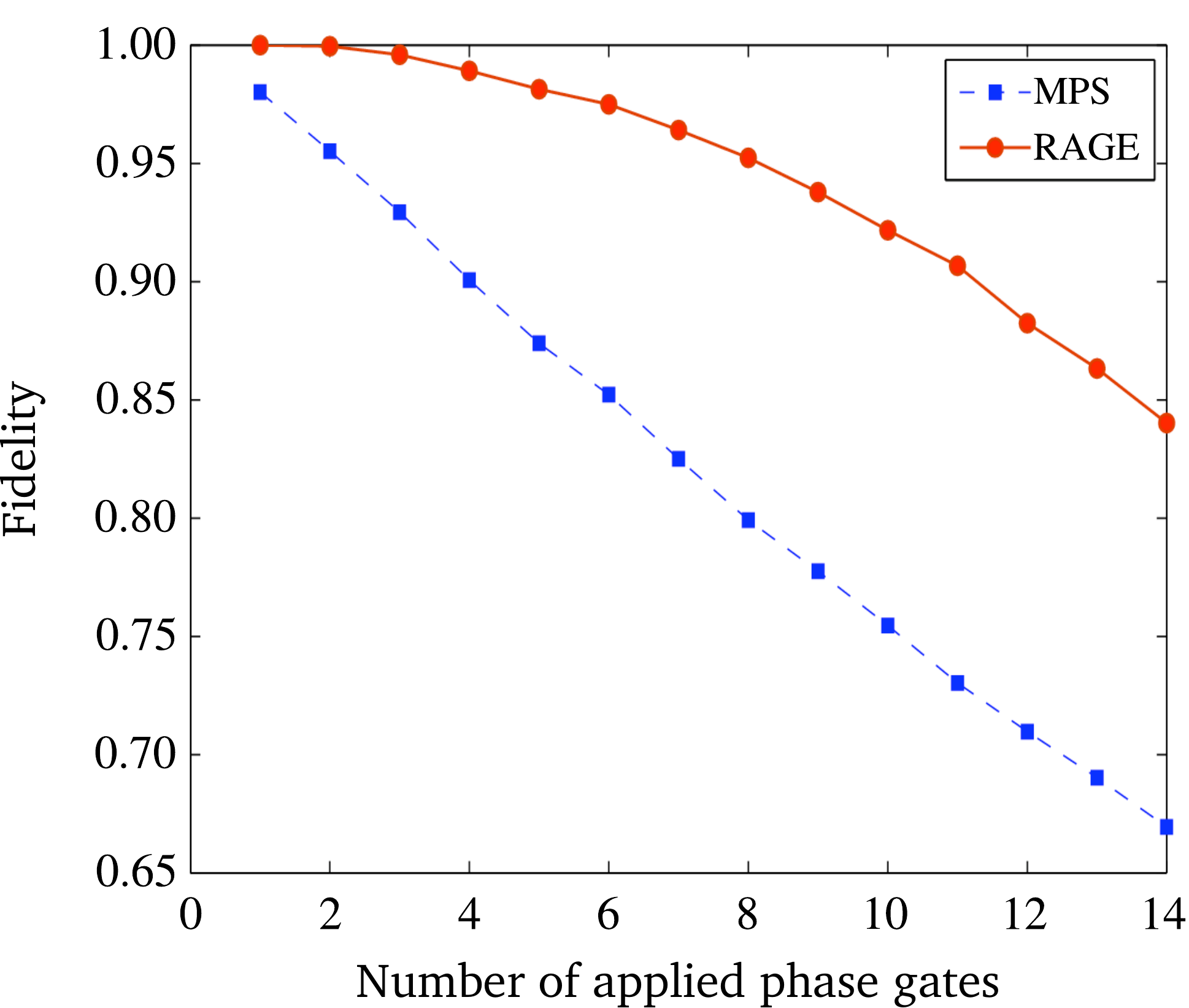}
\caption{(Color online) Comparison of MPS and RAGE with $D = 2$ for the simulation of a random quantum circuit on $N = 14$ qubits. An application of a random local phase gate followed by a random controlled-phase gate with random uniform phase in $[0, 2\pi)$ constitutes one block. For given $k$ we apply this block $k$ times to a randomly chosen initial MPS state. $500$ such runs are determined, and in each the fidelity with the exact state is computed. The average over $500$ realizations is then plotted.}
\label{circuit}
\end{figure}

We have tested the RAGE states in a simulation of a {\it random quantum circuit} (see Fig.~\ref{circuit}) and 
compared the achievable accuracy with MPS. This circuit
has to be applied to an initial state that has an efficient classical description in the first place. We apply 
sequences of (i) random single qubit phase gates and (ii) random controlled phase gates
to an initial random MPS state. In each case, the phase is drawn uniformly from $[0,2\pi)$, as well as the 
support of the gates is drawn from the uniform measure.
For the random MPS as an initial state, random $2\times 2$ matrices have been employed, each of the form
$A+i B$, with the entries of $A$ and $B$ being independently identically distributed on $[-1,1]$. The evolution is then being kept track
of both with a RAGE and an MPS ansatz: In the RAGE ansatz by updating the phases in case of 
the random phase gates and adapting the MPS part in case of local gates, in the MPS case by employing
a sweep in order to find the best MPS ansatz compatible with the new state obtained by the application of a gate
just as being used in DMRG. The plot is obtained from drawing $500$ realizations of the 
resulting stochastic process.
Again, we obtain a significant improvement due to the WGS. 
Here, the advantage of the RAGE ansatz 
is most transparent, as 
the interactions are not local, and it is exactly the long-range nature of the support of the random phase gates that the 
RAGE ansatz can capture well. 
 
Note that similar random circuits -- ones where the phase gates are replaced by random unitaries drawn from the Haar
measure -- give rise to an approximate unitary $3$-design. These can be used to show that for almost any sufficiently long quantum circuit one can construct a black box problem that is
solved by the circuit in a constant number of queries, while requiring exponentially many classical queries, even under
postselection \cite{BH10}. That is to say, a random circuit quite similar to the above one certifies, when setting up a black box problem, 
the superiority of a quantum computer compared to a classical one.

Starting from here, further delineating the 
boundary between efficiently classically simulatable circuits and those exhibiting a speedup offered by 
quantum computation, constitutes an interesting perspective. This topic is also exciting since our results show that the scaling of the 
entanglement entropy alone is not always significant for the classical simulatability of a quantum state. This 
observation has already been made in the context of quantum cellular automata. They give rise to tensor networks 
that are efficiently contractible, yet still slightly violate an area law for the entanglement entropy \cite{BKE10}. 
In the context of random circuits, though, states are encountered that exhibit a much larger degree of entanglement.

\subsection{Conclusions}\label{sec:summary}

\subsubsection{Comparing MPS and TTS}

We have compared MPS and TTS to determine how well they are suited for the description of states relevant in condensed matter physics and to find out where their respective strengths are. We employed both analytical and numerical methods.
Analytically, we have given an example where TTS can offer an advantage for the approximation of states. In the given example, we required a certain entanglement structure to be reflected in the description, and found out that the number of parameters scales more favorably for the TTS than the MPS in this case. This is a motivation, at least under certain circumstances, to include the more general TTS into our RAGE description.

In addition, we have given examples of variational numerical approaches for the description of ground states using TTS and MPS. The used examples are models with interactions which are likely to give rise to entanglement properties that are suspected to be difficult to describe by MPS (long range interactions, broken symmetries.) We found ambivalent results. 

An interesting observation is that although the TTS description does offer advantages, the cases where it is really useful are not easy to identify. The straightforward argument that a possible improvement seems likely in cases where interaction symmetries are broken (being reflected by the broken symmetries of the tree description itself) or where interactions are long-range is too simple. On the other hand, the analytic arguments that can be made lead to a whole class of very suitable states.

\subsubsection{Applications of RAGE states}

We have applied the RAGE states for the simulation of the Ising and the Heisenberg model on two-dimensional lattices, as well as the toric code state (disturbed by local magnetic fields.) The applications of RAGE states leave a two-fold impression. 
On the one hand, we have good results describing the disturbed versions of graph states like the toric code state, using only a very modest amount of parameters. This reflects our intuition that a graph state (having a description in terms of few parameters) under the influence of a local noise model (again: few parameters in the characterization) should be a structure which can be grasped with small amount of data. As we see in the simulations, the disturbance can be well described by the tensor product state combined with the (undisturbed) weighted graph-description. 

However, on the other hand, the RAGE description has its limits. Typical condensed matter systems like the ground state of the 2D Heisenberg model are apparently still to far away from the RAGE states to be approximated well with this set. Apparently, even simple interactions that appear in physical models of condensed matter systems produce very rich ground states, essentially beyond a characterization with weighted phase gates and DMRG methods, even if combined. It is the hope that the present work can contribute to the quest for identifying the set of quantum states
that can be efficiently classically simulated and that in a sense still captures the relevant degrees of freedom of a problem.
\bigskip

\begin{acknowledgments}
We thank F.\ Verstraete, H.\ J.\ Briegel, S.\ Anders, T.\ J.\ Osborne, and
C.\ M.\ Dawson for illuminating discussions and the FWF, the EU (QAP, NAMEQUAM, SCALA, QESSENCE, MINOS, COMPAS), the EPSRC QIP-IRC, the Royal Society, Microsoft Research, the BMBF (QuOReP), the Humboldt Foundation, 
and a EURYI for support. We thank a referee for pointing out interesting references.
Some of the calculations have been carried using facilities of the University of Innsbruck's Konsortium Hochleistungsrechnen.

\end{acknowledgments}


\appendix


\section{The canonical form}\label{sec:orthonormalization}

When optimizing a tensor as in Sec.~\ref{sec:optim in a TTNS}, it is useful to work in a canonical form of the TTS, where in Eq.~\eqref{eq:TTS with T singled out} the set of states $\{\ket{\varphi^1_{\alpha}} \}, \{\ket{\varphi^2_{\beta}} \}, \{\ket{\varphi^3_{\gamma}} \}$ are orthonormal each. In this canonical form, $\tilde{\one}$ from Eq.~\eqref{eq:effective matrices for TTS} is a unit matrix. This spares the calculation of this matrix and moreover simplifies the computation of the energy $E$ and the corresponding tensor $A$, compare also Ref.\ \cite{VDB07}.

Another not immediately obvious benefit of the canonical form is to make the optimization algorithm numerically more stable. The (ordered) spectrum of Schmidt coefficients in the bi-partitions of ground states, which we are searching for in the optimization, shows usually quickly diminishing magnitudes. If the canonical form is \emph{not} used, this property carries over to the quantum mechanical amplitudes used in the definition of the matrix $\tilde{\one}$. Hence, the closer the sweeping procedure gets to the actual ground state, the more singular the matrix $\tilde{\one}$ becomes. This makes the algorithms solving the generalized eigenvalue problem (underlying the minimization problem in TTS unstable.

To define a useful canonical form of a TTS consider the following. Fixating a tensor $A$ in the graph implies a half-order of tensors, defined by the distance of tensors from the tensor $A$ in the tree. Each path in the tree defines a completely ordered set of tensors, which we call a chain of tensors. There is only one leaf in each chain, and it is the last element (the bottom element) of such a chain. Hence there are as many chains as we leaves in the tree. To transform the TTS into its canonical form, we will perform an iterative orthonormalization procedure to each chain. We start from the bottom, which is a leave node and go to the top, which is a tensor connected to the tensor $A$ to be optimized.

As the leaves of the tree correspond to single particle spin systems, we can easily find an orthonormal basis for this system. Beginning with the leaves we will from now on assume all subsystems in a chain are already orthonormal.
\begin{figure}[t]
\includegraphics[width=0.55\columnwidth]{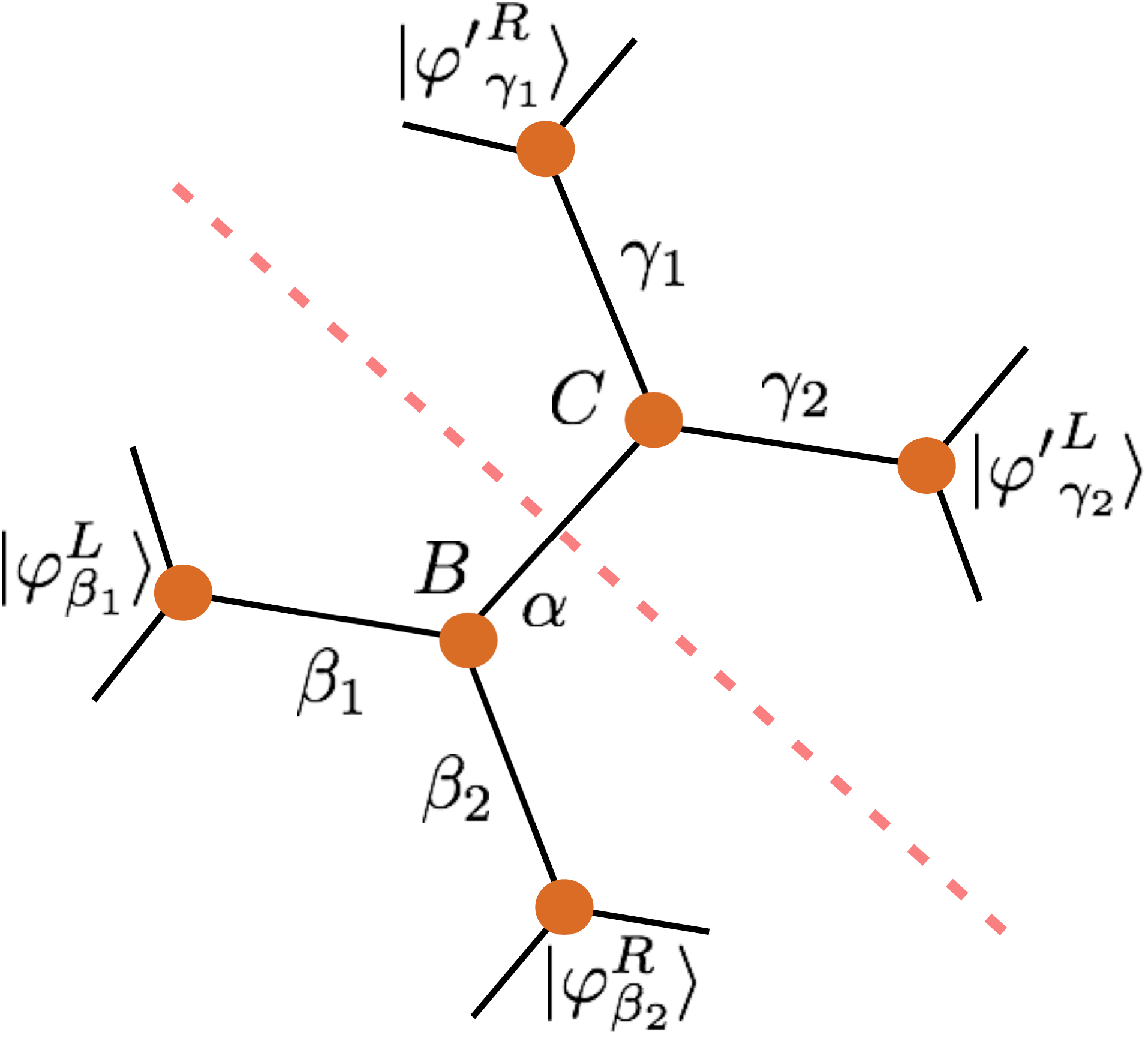}
\caption{(Color online) A bi-partition of a TTS along an edge, leading to two simply connected sub states. The Schmidt decomposition can be inferred directly from the tensors adjacent to the edge.}
\label{fig:cutting edge}
\end{figure}
Consider the situation in Fig.~\ref{fig:cutting edge} and let tensor $C$ be the lower one in the chain, \ie the position of tensor $C$ in the chain is $n$ and the position of tensor $B$ in the chain is $n+1$. With the two lower-level subsystems $\{ \ket{\varphi^1_{\beta}(n-1)} \}$ and $\{ \ket{\varphi^2_{\gamma}(n-1)} \}$ stemming from the two chains spawned from tensor $C$ when going away from tensor $A$, we formulate the state belonging to the sub state linked to the tree via edge $\alpha$
\be
\ket{\varphi^1_{\alpha}(n)} = \sum_{\beta_1, \beta_2} B_{\alpha ,\beta_1, \beta_2} \ket{\varphi^1_{\beta_1}(n-1)}\ket{\varphi^2_{\beta_2}(n-1)}.
\label{eq:phin}
\ee
The orthonormality condition for the vectors $\{\ket{\varphi_{\alpha}(n)}\}$ reads
\bee
\begin{split}
\delta_{\alpha',\alpha} =& \braket{\varphi^1_{\alpha'}(n)}{\varphi^1_{\alpha}(n)}\\
=&\sum_{\beta'_1 ,\beta'_2 ,\beta_1, \beta_2} B^*_{\alpha' ,\beta'_1 ,\beta'_2} B_{\alpha ,\beta_1 ,\beta_2}\\
&\quad \times \braket{\varphi^1_{\beta'_1}(n-1)}{\varphi^1_{\beta_1}(n-1)} \braket{\varphi^2_{\beta'_2}(n-1)}{\varphi^2_{\beta_2}(n-1)}\\
=&\sum_{\beta_1 ,\beta_2} B^*_{\alpha'  , \beta_1 ,\beta_2} B_{\alpha ,\beta_1, \beta_2}.
\label{eq:deltarelation}
\end{split}
\eee
With index reordering $\beta_1 ,\beta_2 = (\beta_1, \beta_2)$, these tensors are matrices and we can apply a QR-decomposition
\be
B_{\alpha (\beta_1 ,\beta_2)} = \sum_{\sigma} Q_{(\beta_1, \beta_2) \sigma} R_{\sigma, \alpha}
\ee
with an orthonormal matrix $Q$ and an upper triangular matrix $R$. As the state vectors $\ket{\varphi^1_{\alpha}(n)}$ themselves are 
correspond to subsystems of a state higher in the chain, we can absorb the matrix $R$ into the tensor of the higher level, in this case
\be
\ket{\varphi^1_{\gamma_2}(n+1)}= \sum_{\alpha, \gamma_1} D_{\gamma_2 ,\alpha ,\gamma_1} \ket{\varphi^1_{\alpha}(n)}\ket{\varphi^2_{\gamma_1}(n)}
\label{eq:phinplusone}
\ee
and we can absorb $R$ into $D$
\be
D_{\gamma_2 ,\alpha ,\gamma_1} \mapsto D'_{\gamma_2 ,\alpha ,\gamma_1} = \sum_{\alpha'} R_{\alpha,\alpha'} D_{\gamma_2 ,\alpha' ,\gamma_1}
\ee
while at the same time replacing $B$ by $Q$ (with reshaped indices.) This makes Eq.~(\ref{eq:deltarelation}) true. As a QR-decomposition is an efficient algorithm (and here applied to matrices with dimension $D \times D^2$), each step is efficient. All in all, there are $N$ chains to consider and maximally $N$ tensors in a chain, making the treatment of the whole tree efficient. This orthonormalization procedure is consistent between different chains, as there are different elements in different chains, but no two chains contain the same two tensors in different order. The last $R$ matrix of any chain, which cannot be absorbed anymore, can be thrown away, as it would be absorbed into the tensor to be found in the eigenvalue procedure anyway.

This procedure is applicable \emph{as is} to open-boundary MPS, as they are a subset of the TTS. Unfortunately, closed-boundary MPS suffer from similar numerical instabilities when approaching the ground state. In this case, we apply the procedure given above to a version of the closed-boundary MPS which is made open-boundary by cutting an arbitrary edge. Beware, in this case the matrix $\tilde{\one}$ is not a unit matrix, so we still have to solve a generalized eigenvalue problem, but at least $\tilde{\one}$ is not close to singular anymore, providing numerical stability.

\section{Projected entangled pair states}\label{sec:PEPS}

We want to point the readers' attention to the fact that the basic procedure, underlying all evaluations of observables and also the variational methods in tensor-network states, is \emph{the summation over all possible values} of the indices of the tensors in the description of the tensor-network state. The presented algorithms of Secs.~\ref{sec:observable evaluation in MPS} and \ref{sec:observable evaluation in TTN} circumvent the complexity of the contraction problem exploiting additional properties of the respective tensor-networks. If a generic network has such exploitable favorable properties is in many cases unknown. 

In the case of MPS and TTS, in particular, we make use of the essentially one-dimensional structure of the graphs underlying the network, \eg by using redundancy in the recursive procedure of a tree-contraction, or the possibility to write down a sum of many matrix products in shorter form using, essentially, the distributive law. However, even the possibility to efficiently contract a network alone is not a sufficient criterion for the possibility to efficiently approximate ground states. Even if each step in an updating procedure can be performed efficiently, the convergence of the procedure over many steps can be very slow. It is \eg possible to encode computationally hard problem as a ground state approximation problem in an MPS, as has been shown in Ref.~\cite{SCV08}.

Another type of state \emph{not} corresponding to a linear structure is the projected entangled pair state (PEPS) \cite{VC06b}. The PEPS are tensor-network states, similar to the MPS and TTS, but correspond to graphs with loops. This generalization has major consequences, improving the entanglement properties on the one hand, but making it necessary to find new algorithms for computing expectation values and updating the description on the other. There are no known efficient algorithms for the \emph{exact} contraction of such a network. This might be due not to a lack of ideas but an inherent computational hardness of the problem \cite{SWVC07}. 
Approximate contractions are still possible, as will be shown.

\subsection{Definition and properties}

\begin{figure}[t]
\includegraphics[width=0.7\columnwidth]{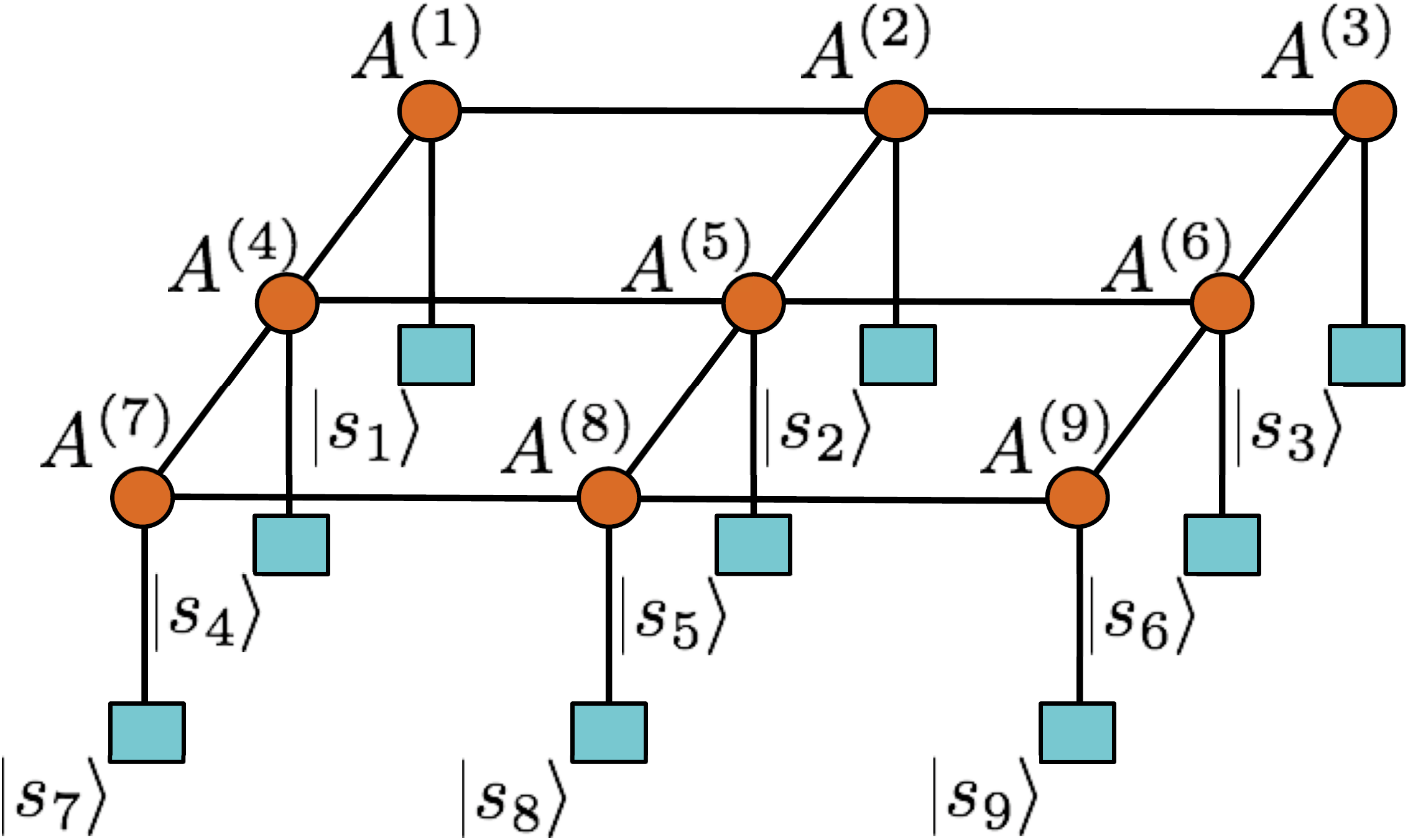}
\caption{(Color online) The graph corresponding to a two-dimensional PEPS shows a higher connectivity than the graph corresponding to an MPS.}
\label{fig:PEPS_structure}
\end{figure}

It is not necessary to restrict the number of unphysical indices per tensor to two, like in the example of MPS, but in principle any number is possible. Consider, for example, the following network structure
\begin{multline}
A_{s_1,\ldots, s_9}= \sum_{\text{Greek indices}} A_{\alpha_1,\beta_1}^{(1)s_1}
A_{\alpha_1,\alpha_2,\beta_2}^{(2)s_2}
A_{\alpha_2,\beta_3}^{(3)s_3}
A_{\beta_1,\alpha_3,\beta_4}^{(4)s_4}\\
A_{\alpha_3,\beta_2,\alpha_4,\beta_5}^{(5)s_5}
A_{\beta_3,\alpha_4,\beta_6}^{(6)s_6}
A_{\beta_4,\alpha_5}^{(7)s_7}
A_{\alpha_5,\beta_5,\alpha_6}^{(8)s_8}
A_{\alpha_6,\beta_6}^{(9)s_9}.
\end{multline}
Here, the tensor with components $A_{s_1,\ldots, s_9}$, connecting to the physical indices, and used to define the PEPS
\be
\ket{\psi_{\mathrm{PEPS}}}=\sum_{s_1,\ldots ,s_9} A_{s_1,\ldots ,s_9} \ket{s_1,\ldots ,s_9}
\ee
is obtained by summing over a network of nine tensors, which are connected to the nearest neighbor in a two-dimensional grid. The network corresponds to the graph depicted in Fig.~\ref{fig:PEPS_structure}.

To understand the entanglement properties of PEPS, it is useful to give the MPS description another interpretation, which can be found in the construction and analysis of the famous AKLT-model~\cite{AKLT88}. In this context, a subsystem of two contiguous sites $\ket{s_n, s_{n+1}}$, as taken from \eg Eqs.~\eqref{eq:MPS CBC} and \eqref{eq:MPS OBC}, and corresponding to two tensors (or matrices) $A^{(n)s_{n}},A^{(n+1)s_{n+1}}$,
\be
\sum_{\beta} A^{(n)s_{n}}_{\alpha, \beta} A^{(n+1)s_{n+1}}_{\beta ,\gamma} \ket{s_n ,s_{n+1}}
\ee
is interpreted as a \emph{projection} of a maximally entangled pair of auxiliary sites, $\sum_{\beta=0}^{D-1} \ket{\beta}\ket{\beta}$, using the tensors $A^{(n)s_{n}},A^{(n+1)s_{n+1}}$ as projectors.

The upper bound on the bi-partite entanglement of an MPS, as implied by cutting edge $\beta$, is directly related to the number of maximally entangled pairs which are cut in the bi-partition. This picture very naturally carries over to the PEPS construction. Depending on the connectivity of sites, PEPS can fulfill area-- and even volume laws.

\subsection{Evaluation of observables and variational methods}\label{sec:PEPSeval}

In this subsection we will briefly outline how to evaluate an observable in a PEPS and how to apply variational methods to a PEPS, which can be based on evaluations of expectation values and overlaps as well. More precisely, similar to the case of TTS and MPS, the computation of an expectation value in a PEPS amounts to a contraction, \ie a summation over the indices, of the tensor-network, see also Sec.~\ref{sec:PEPS}. As this is a computationally hard problem in principle, we want to make use of an inherent structure in the description of the PEPS in order to simplify the contraction problem, similar to the case of MPS and TTS. 

The structure to be exploited is the equivalence of a) contracting a $d$-dimensional PEPS to b) the transfer of a $(d-1)$-dimensional quantum system. As an example, consider the computation of the norm of a two-dimensional rectangular PEPS, corresponding to a graph as given in Fig.~\ref{fig:PEPS_structure}. The treatment of the transfer problem is in this case simplified by the fact that the one-dimensional quantum system is given by an MPS and the transfer operators are given by matrix product operators (MPO) applied to the MPS. This problem is can be treated effectively using \emph{approximate} contraction methods. The algorithm we refer to in the following was introduced in Ref.~\cite{VC06b}.

The expression to be computed, serving as an example is the square of the norm $\braket{\psi_{\mathrm{PEPS}}}{\psi_{\mathrm{PEPS}}}$,
\begin{figure}[t]
\includegraphics[width=0.7\columnwidth]{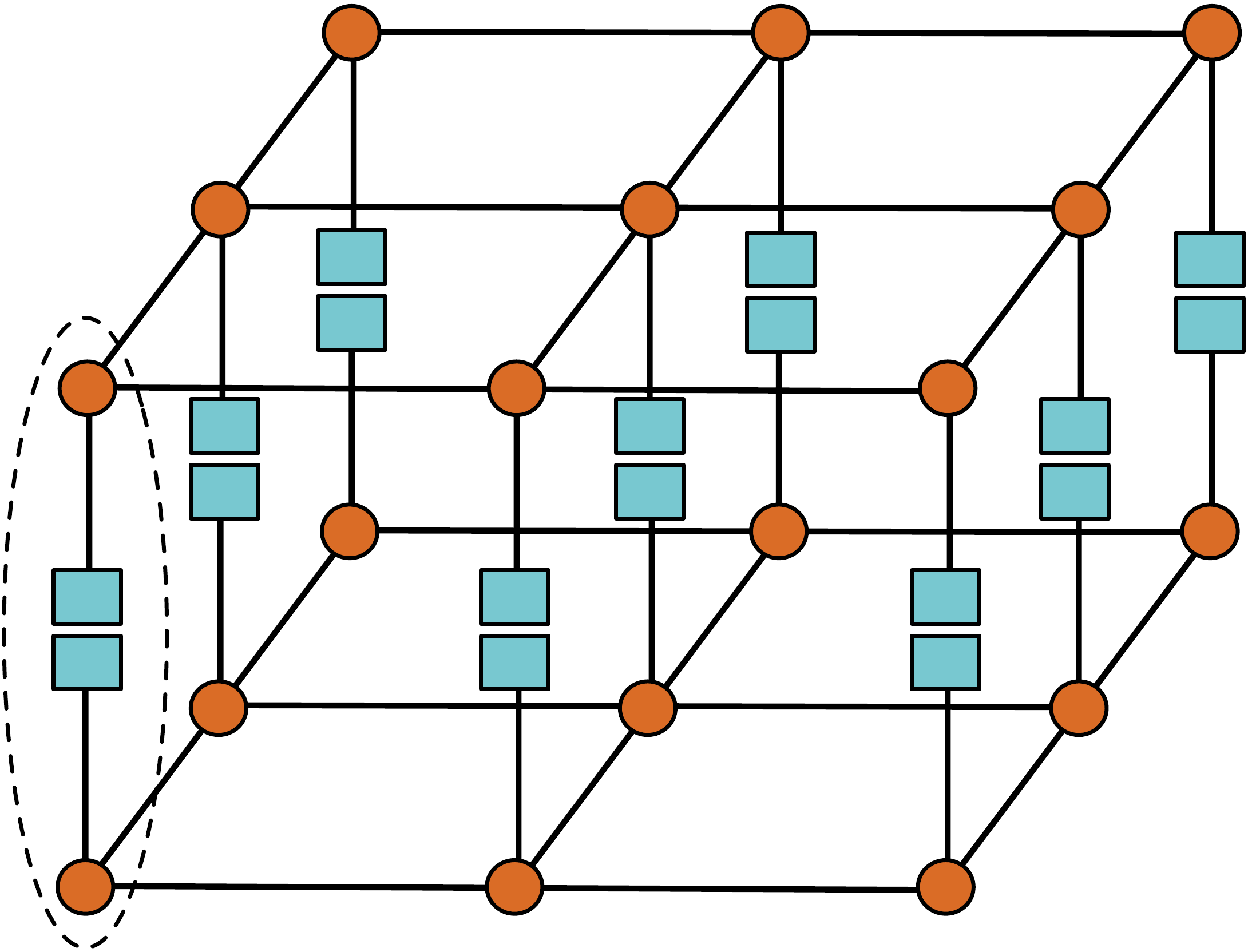}
\caption{(Color online) Schematic representation of a PEPS contraction. The top layer corresponds to the ket-PEPS and the lower layer corresponds to the bra-PEPS. The dashed ellipse denotes the tensor structure which is contracted first, resulting in a single-layered rectangular tensor grid. The resulting rectangular grid is then contracted using MPS methods.}
\label{fig:PEPS_contraction}
\end{figure}
and corresponds to Fig.~\ref{fig:PEPS_contraction}. The top layer, as depicted in the figure, corresponds to the ket-PEPS and the lower layer corresponds to the bra-PEPS. To reduce the tensor grid to the transport problem in a one-dimensional quantum system, we first contract the physical sites, \ie we perform the contraction $\braket{s'_i}{s_i}$ for all values of $s,s'$ and $i$, leading to new tensors, \eg
\begin{multline}
\sum_{s_5,s'_5} A_{\alpha_3,\beta_2,\alpha_4,\beta_5}^{(5)s_5}\braket{s'_5}{s_5} A_{\alpha'_3,\beta'_2,\alpha'_4,\beta'_5}^{*(5)s'_5} \mapsto\\
 B_{(\alpha_3 ,\alpha'_3) (\beta_2 ,\beta'_2) (\alpha_4 ,\alpha'_4) (\beta_5 ,\beta'_5)}^{(5)} = \sum_{s_5}A_{\alpha_3,\beta_2,\alpha_4,\beta_5}^{(5)s_5} 
A_{\alpha'_3,\beta'_2,\alpha'_4,\beta'_5}^{*(5)s_5}.
\label{eq:Bdef}
\end{multline}
Here, $B$ is a tensor of rank $4$ with new effective indices $(\alpha_3 ,\alpha'_3)$, etc. This procedure results in a rectangular grid. The computation of product observables is similar, the expressions $\braket{s'_i}{s_i}$ being replaced by $\bra{s'_i}\mathcal{O}^{(i)}\ket{s_i}$ with a local observable $\mathcal{O}^{(i)}$ acting on site $i$. For the updates in a variational method, a replacement of tensors with tensors of type $D$ in Eq.~\eqref{eq:dummytensor} can be performed in a straightforward manner, leading to quadratic forms analogously to the case of MPS.

We treat the grid of $B$-tensors as follows. The first (\eg horizontal) line of tensors at the boundary can be interpreted as an MPS, where the lower indices are considered open. The second line can be viewed as an MPO acting on the first MPS. The resulting state (after contracting two lines) can again be described by an MPS, but of increased dimension. The aim is now to find (\eg via a variational method) the optimal approximation of the resulting state by an MPS of {\em fixed} (low) dimension. This is \eg done by optimizing the individual tensors via solving a generalized eigenvalue problem. The MPS found this way is now processed further, \ie the MPO corresponding to the third line of tensors is applied, and one again aims at obtaining a proper approximation of the resulting state by an MPS of fixed dimension. The process is repeated until the second to last line of tensors is reached. The final step corresponds to calculating the overlap of the MPS resulting from above procedure (after processing all but the final line), and the MPS corresponding to the final line. Summarizing: (i) Start with $i=1$ and set $\langle\tilde{M}_1| := \langle M_1|$. (ii) Apply the MPO $M_{i+1}$ to the intermediate MPS $\langle\tilde{M}_{1,\ldots,i}|$. Both having a small bond-dimension, we obtain an MPS of large bond-dimension, $\langle M_{1,\ldots,i+1}|$. (iii) Reduce the bond-dimension of $\langle M_{1,\ldots,i+1}|$ to obtain another intermediate MPS $\langle \tilde{M}_{1,\ldots,i+1}|$, representing $\langle M_{1,\ldots,i+1}|$ as good as possible with this smaller bond-dimension. (iv) Increase $i$ by one and continue with step (ii). For details of the method, we refer 
the reader to Ref.\ \cite{MVC07}. The precision of this kind of 
contraction can moreover be improved by an error-correction scheme, introduced in Ref.\ \cite{HND09}.

The accuracy of the approximation depends on the \emph{compressibility} of the intermediate MPS with large dimension into a version with smaller dimension. This compressibility is hence related to the amount of entanglement in the intermediate MPS which is created by the applied MPO. A PEPS that is close to a product state clearly induces MPS and MPO that produce intermediate MPS with comparatively little entanglement., and hence can be treated well by the given method.

\section{RAGE states based on PEPS}

As shown in Sec.~\ref{sec:PEPS}, in a PEPS any (physical) site of a chosen subset of an arbitrary subsystem can share one (or more) maximally entangled pairs of auxiliary sites with any (physical) site outside the subsystem. Hence the upper bound on the block-wise entanglement in the state can not only be increased by increasing the index dimension $D$ of the tensors (corresponding to the dimensionality of the auxiliary system), but also by raising the number of pairs shared between the subsystem and the remainder of the system. This is the main difference to the MPS and TTS described above. Another important observation is that even if we restrict the sharing of auxiliary pairs to nearest neighbors, a volume law can be satisfied. This makes the PEPS description very powerful.
We will now consider the combination of PEPS and WGS. Although PEPS can fulfill a volume law in any dimension, it is in some cases advantageous to put some information from the PEPS description into a WGS.

\subsection{Why to apply WGS to PEPS}

As shown in Sec.~\ref{sec:PEPSeval}, the method to optimize a PEPS tensor grid follows an iterative procedure, where tensors will be updated individually one by one while keeping the others fixed, and repeating the step for all tensors in the grid several times. In such a procedure it is useful to start with a good first guess for the tensor entries. Generically such a good starting point is not found easily, but there are exceptions. Moreover, the updating procedure is based on contractions of the grid, which scales rather unfavorably (though by definition efficiently) with $D^{12}$. Hence, small values of $D$ are desirable.

Let us reconsider now the example of a description of a noisy (\ie slightly disturbed) graph or stabilizer state. The undisturbed graph state can be represented by the WGS alone, leaving the noise description to the tensorial part of the RAGE state. Noise has in many cases the property to be uncorrelated over distances and hence the tensorial part of the RAGE state corresponds to an uncorrelated state. This kind of state is usually close to a product state and can be described well by low dimensional MPS/MPO or PEPS. This setting is thus ideal for a RAGE description based on PEPS with a small value of $D$ and precalculated phases for the WGS part. The initial state for the PEPS is then additionally chosen to be a product state, providing a good first guess.

\subsection{How to apply WGS to PEPS}

The basis of the efficient treatment of RAGE states is the efficient calculation of reduced density matrices with small support. As an example, we will give the formulas for the reduced density operator on a system of two qubits $\mathcal{S}=\{a,b\}$ with complement $\bar{\mathcal{S}}$, which can be generalized in a straightforward fashion. Similar to Sec.~\ref{sec:TTSWGScomp} we use the operators $W(\varphi)$, $\mathcal{W}_{\varphi}$ and $\mathcal{W}_{a,b}$ as defined in Eqs.~\eqref{eq:Wdef}, \eqref{eq:WWdef} and \eqref{eq:WWabdef}.

We obtain the following formula for the elements of the reduced density matrix
\bee
\begin{split}
\bra{t'_a, t'_b}\rho_{\mathcal{S}} \ket{t_a ,t_b}&= \bra{\psi_{\text{PEPS}}}\mathcal{W}_{\varphi}^{\dagger} \ketbra{t_a}{t'_a}^{(a)}\ketbra{t_b}{t'_b}^{(b)} \mathcal{W}_{\varphi} \ket{\psi_{\text{PEPS}}}\\
&=\sum_{{s_1,\dots ,s_N}\atop{s'_1,\dots ,s'_N}} \bra{\mathbf{s}} \mathcal{W}_{\varphi}^{\dagger} \ketbra{t_a}{t'_a}^{(a)}\ketbra{t_b}{t'_b}^{(b)} \mathcal{W}_{\varphi} \ket{\mathbf{s}'} \\
&\quad \times \sum_{\text{Greek indices}} A^{(1)*}_{\alpha ,\beta, \gamma ,\delta, s_1}  \dots  A^{(1)}_{\alpha' ,\beta' ,\gamma' ,\delta' ,s'_1}\\
&=\sum_{{s_1,\dots ,s_N}\atop{s'_1,\dots ,s'_N}} \bra{\mathbf{s}} \mathcal{W}_{a,b}^{\dagger} \ketbra{t_a}{t'_a}^{(a)}\ketbra{t_b}{t'_b}^{(b)} \mathcal{W}_{a,b} \ket{\mathbf{s}'} \\
&\quad \times \sum_{\text{Greek indices}} A^{(1)*}_{\alpha ,\beta ,\gamma ,\delta ,s_1}  \dots  A^{(1)}_{\alpha' ,\beta' ,\gamma' ,\delta' ,s'_1}.
\end{split}
\eee
The application of the operators $\mathcal{W}_{a,b}$ and the subsequent 
contraction of the bra-ket terms leaves us with tensors
\be
B_{(\alpha, \alpha') (\beta, \beta') (\gamma ,\gamma') (\delta, \delta')}^{(k)} = \sum_{s_k} A_{\alpha,\beta,\gamma,\delta}^{(k)s_k} A_{\alpha',\beta',\gamma',\delta'}^{*[k]s_k},
\ee
as known from the original PEPS contraction ansatz, but now modified by 
phases stemming from the terms
\bee
\bra{\mathbf{s}} \mathcal{W}_{a,b}^{\dagger} \ketbra{t_a}{t'_a}^{(a)}\ketbra{t_b}{t'_b}^{(b)} \mathcal{W}_{a,b} \ket{\mathbf{s}'},
\eee
similar to the situation in RAGE states based on MPS and TTS, where the initial tensor grid is modified by phases as well. The so modified grid resulting from the local contractions can then be contracted using the algorithm given in Sec.~\ref{sec:PEPSeval}.

\end{document}